\documentclass[12pt,a4paper]{article}

\usepackage[T1]{fontenc}
\usepackage[latin1]{inputenc}
\usepackage[nosort]{cite}
\usepackage{color}
\usepackage[bulletsep]{collref}
\usepackage{graphicx}
\usepackage{bbm}
\usepackage{amsmath}
\usepackage{amssymb}
\usepackage{physics}
\usepackage{subfig}
\usepackage{verbatim}
\usepackage{multirow}
\usepackage{tikz}
\usepackage{amsthm}
\usepackage{fancyhdr}
\usepackage{wrapfig}
\usepackage{hyperref}
\usepackage{dsfont}
\usepackage{shuffle} 
\usepackage{mathtools}

\makeatletter \@addtoreset{equation}{section} \makeatother

\makeatletter
\let\old@startsection=\@startsection
\let\oldl@section=\l@section
\renewcommand{\@startsection}[6]{\old@startsection{#1}{#2}{#3}{#4}{#5}{#6\mathversion{bold}}}
\renewcommand{\l@section}[2]{\oldl@section{\mathversion{bold}#1}{#2}}
\makeatother

\makeatletter
\let\old@makecaption=\@makecaption
\def\@makecaption{\small\old@makecaption}
\makeatother

\newcommand{\bea}{\begin{eqnarray}}
\newcommand{\eea}{\end{eqnarray}}
\newcommand{\be}{\begin{eqnarray}}
\newcommand{\ee}{\end{eqnarray}}
\newcommand{\ben}{\begin{eqnarray*}}
\newcommand{\een}{\end{eqnarray*}}
\newcommand{\beq}{\begin{equation}}
\newcommand{\eeq}{\end{equation}}

\makeatletter
\def\mr@ignsp#1 {\ifx\:#1\@empty\else #1\expandafter\mr@ignsp\fi}%
\newcommand{\multiref}[1]{\begingroup
\xdef\mr@no@sparg{\expandafter\mr@ignsp#1 \: }%
\def\mr@comma{}%
\@for\mr@refs:=\mr@no@sparg\do{\mr@comma\def\mr@comma{,}\ref{\mr@refs}}%
\endgroup}
\makeatother

\newcommand{\hypref}[2]{\ifx\href\asklfhas #2\else\href{#1}{#2}\fi}

\newcommand{\appref}[1]{appendix~\multiref{#1}}

\renewcommand{\eqref}[1]{(\multiref{#1})}

\ifx\href\asklfhas\newcommand{\href}[2]{#2}\fi

\usepackage{color}

\begin{document}

\thispagestyle{empty}

\begin{flushright}\footnotesize
\texttt{HU-Mathematik-2021-06}\\
\texttt{HU-EP-21/51}\\
\texttt{SAGEX-21-37-E}
\end{flushright}
\vspace{1cm}
\begin{center}%
{\LARGE\textbf{\mathversion{bold}%
Combinatorial Solution of the \\ \vspace{0.2cm}
Eclectic Spin Chain}}

\vspace{1cm}
 \textsc{Changrim Ahn$^a$, Luke Corcoran$^b$, Matthias Staudacher$^{a,b}$} \vspace{8mm} 
 \\
\textit{%
$^a$ Department of Physics, Ewha Womans University, \\
52 Ewhayeodae-gil, Seodaemun-gu, Seoul 03760, S. Korea
}
 \\ \vspace{0.5cm}
\textit{%
$^b$ Institut f\"{u}r Physik, Humboldt-Universit\"{a}t zu Berlin, \\
Zum Gro{\ss}en Windkanal 2, 12489 Berlin, Germany
}

\texttt{\\ ahn@ewha.ac.kr,\{corcoran,staudacher\}@physik.hu-berlin.de}
%

\par\vspace{25mm}

\textbf{Abstract} \vspace{5mm}

\begin{minipage}{13cm}
The one-loop dilatation operator in the holomorphic 3-scalar sector 
of the dynamical fishnet theory is studied. Due to the non-unitary nature of the underlying field theory this operator, dubbed in \cite{Ipsen_2019} the eclectic spin chain Hamiltonian, is non-diagonalisable. The corresponding spectrum of Jordan blocks leads to logarithms in the two-point functions, which is characteristic of logarithmic conformal field theories. It was conjectured in \cite{Ahn_2021} that for certain filling conditions and generic couplings the spectrum of the eclectic model is equivalent to the spectrum of a simpler model, the hypereclectic spin chain. We provide further evidence for this conjecture, and introduce a generating function which fully characterises the Jordan block spectrum of the simplified model. This function is found by purely combinatorial means and is simply related to the $q$-binomial coefficient.
\end{minipage}
\end{center}
\newpage
\tableofcontents
\medskip
\hrule

%
%
%


\section{Introduction and Overview}
Integrability of gauge and string theories continues to generate exciting new types of exactly solvable models, ranging from new spin chains to novel quantum field theories and hitherto unstudied string theories. Frequently this stems from deformations and/or subtle limit taking of previously investigated systems. A case at hand is a certain double-scaling limit of the three-parameter $\gamma$-deformation of $\mathcal{N}=4$ super Yang-Mills theory that had originally been proposed in \cite{Frolov_2005,Frolov2_2005}. It partially or fully breaks R-symmetry and thus also supersymmetry, while apparently retaining conformality and integrability in the planar limit. 
Its double-scaling limit was then proposed in \cite{Gurdogan_2016}, combining a strong imaginary $\gamma$-twist with a vanishing coupling constant. In this limit all gauge field interactions decouple and one is left with an un-gauged quantum field theory of scalars $\phi_i$ and fermions $\psi_i$, $i=1,2,3$. In light of \cite{Kazakov_2019} we refer to the resulting theory as the \textit{dynamical fishnet theory}, with interaction Lagrangian 
\begin{align}\label{eq:Ldfn}
        \mathcal{L}^{\text{int}}_{\text{DFN}}&=N_c \text{tr}\left(\xi_1^2\phi^{\dagger}_2\phi^{\dagger}_3\phi_2\phi_3+\xi_2^2\phi^{\dagger}_3\phi^{\dagger}_1\phi_3\phi_1+\xi_3^2\phi^{\dagger}_1\phi^{\dagger}_2\phi_1\phi_2\right)\\
        &+N_c\text{tr}\left(i\sqrt{\xi_2\xi_3}(\psi^3\phi_1\psi^2+\bar{\psi}_3\phi_1^\dagger \bar{\psi}_2)+\text{cyclic}\right)\notag.
\end{align}
This model can be further simplified by taking $\xi_1=\xi_2=0, \xi_3\equiv\xi$. In this case we recover the bi-scalar \textit{fishnet theory}
\begin{equation}\label{eq:Lfn}
\mathcal{L}_{\text{FN}}^{\text{int}}=\xi^2N_c \text{tr}\left(\phi^{\dagger}_1\phi^{\dagger}_2\phi_1\phi_2\right).
\end{equation}
Notably, the theories \eqref{eq:Ldfn} and \eqref{eq:Lfn} are non-unitary. However, the chiral nature of these interactions leads to a vast simplification in the Feynman-diagrammatic structure of many physical quantities. This raises the hope that the integrability of these models might be more easily understood from first principles. Recall that the origin of integrability of undeformed or $\gamma$-deformed $\mathcal{N}=4$ SYM remains shrouded in mystery. On the contrary, in the chiral models one often observes recursive structures in the Feynman graphs, and the associated graph-building operator may sometimes be shown to possess integrable properties \cite{Kazakov_2019,Gromov_2018,Gromov3_2019}. In some cases, correlation functions are represented by a single Feynman diagram. An example of this is the fishnet Feynman integrals. These have been shown to enjoy a Yangian symmetry \cite{Chicherin_2017}, which in some cases has been sufficient to bootstrap the integral \cite{Loebbert_2020,Corcoran_2021}. In a four-point limit these fishnet graphs reduce to the celebrated Basso-Dixon correlators, for which integrability has been studied from various perspectives \cite{Basso_2017, Derkachov_2020,Basso_2021}. The fishnet theory has also been argued to possess at strong coupling a holographic dual \cite{Gromov_2019,Gromov2_2019,Gromov_2020}.

As written, the theories \eqref{eq:Ldfn} and \eqref{eq:Lfn} are not strictly conformal, even in their planar limit \cite{Fokken_2014,Sieg_2016}. Double trace couplings are generated upon renormalisation. However, it has been argued in \cite{Kazakov_2019,Grabner_2018} that these coupling may be fine-tuned as a function of $\xi^2$ such that the overall beta-function becomes identically zero, while preserving integrability. As a result, one expects to get an integrable {\it logarithmic} conformal field theory. This is a consequence of the models' non-unitarity: While the state space is still reducible, it is no longer decomposable. The logarithmic nature of the underlying CFT poses curious new challenges for the spectral problem of the theory \cite{Ipsen_2019,Ahn_2021}. In particular, in certain operator sectors the dilatation operator is no longer diagonalisable. It is known that this leads to the appearance of logarithms in the two-point functions \cite{Hogervorst_2017}. For example, in the simplest case where the dilatation operator acts on an operator pair $\mathcal{O}_1, \mathcal{O}_2$ as a $2\times 2$ Jordan cell
\begin{equation}
\mathfrak{D}\begin{pmatrix}\mathcal{O}_1\\\mathcal{O}_2\end{pmatrix}= \begin{pmatrix}
    \Delta & 1\\0&\Delta
    \end{pmatrix}\begin{pmatrix}\mathcal{O}_1\\\mathcal{O}_2\end{pmatrix},
\end{equation}
the two-point function can be brought into the form
\begin{equation}
\langle \mathcal{O}_i(x)\mathcal{O}_j(0)\rangle=\frac{c}{|x|^{2\Delta}}\begin{pmatrix}\log x^2&1\\1&0\end{pmatrix}.
\end{equation}
An explicit example of this in the fishnet theory for length $5$ operators is given in \cite{Gromov_2018}.

Logarithmic conformal field theories play an important role in two dimensions \cite{1993log}. There, due to their direct connection with two-dimensional statistical mechanics models, they are of great physical interest. Important examples include models of self-avoiding walks, polymers, and percolation; for recent progress see \cite{he2021note} and references therein. Often the logarithmic scaling violations occurring in these models are of both experimental and theoretical  interest. In fact, their mathematical analysis often shows intricate and novel features as compared to the non-logarithmic case. In higher dimensions, logarithmic CFTs have been much less studied. Still, given their success story in two dimensions, it is natural to suspect that they will also be of considerable value.

A systematic study of the dilatation operator in strongly twisted $\mathcal{N}=4$ SYM was initiated in \cite{Ipsen_2019}. It was found that the mentioned non-diagonalisability is ubiquitous in these models, leading to a rich structure of Jordan cells. It was also pointed out that the standard methods of integrability largely fail when applied to the model's non-diagonalisable sectors. This was then studied in more detail in a particularly simple setting, namely at one-loop and with three scalars of equal chirality, in \cite{Ahn_2021}. The resulting spin chain was dubbed the \textit{eclectic spin chain} in \cite{Ipsen_2019,Ahn_2021}, and an even simpler model, the \textit{hypereclectic spin chain} was proposed, but not solved. Interestingly, the latter appeared to possess an even richer spectrum of Jordan decompositions as compared to the one in the generic eclectic model. This phenomenon was called \textit{universality} in \cite{Ahn_2021}.

The current work seamlessly continues \cite{Ahn_2021}, and proceeds to find the exact solution of the hypereclectic model. Curiously, for the moment this does not use at all the model's integrability, but instead combines methods of linear algebra and combinatorics. As a result we obtained an elegant generating function for the spectrum of Jordan blocks. It is reminiscent of a partition function, since it can be obtained by computing a trace over the state space
\begin{equation}\label{eq:Z1intro}
\mathcal{Z}(q)=\text{tr} \hspace{0.05cm}q^{\hat{S}'}\,,
\end{equation}
where $\hat{S}'$ is a certain counting operator, which is diagonal in the canonical basis of tensor product states of the spin chain, see end of section \ref{sec:generalLMK}.
It uniquely encodes in full generality the sizes and multiplicities of the Hamiltonian's Jordan block decomposition:
\begin{align}\label{eq:Z2intro}
&\mathcal{Z}(q)=\sum_{j} N_j [j]_q=N_1\, q^0 + N_2\, \left( q^{-\frac{1}{2}}+q^{\frac{1}{2}} \right) 
+ N_3\, \left( q^{-1}+q^0+q^1 \right) +\ldots \,,
\end{align}
where $N_j$ is the number of Jordan blocks of length $j$, and $[j]_q$ is a $q$-analog of $j$, cf.\ \eqref{eq:qnumber}.
It is easy to see that the $\{N_j\}$ are indeed uniquely fixed once one knows $\mathcal{Z}(q)$.
We also derive formulas expressing $\mathcal{Z}(q)$ more explicitly than \eqref{eq:Z1intro} in terms of $q$-binomial coefficients. For example, for the case corresponding to the fishnet interaction Lagrangian \eqref{eq:Lfn}, with $L-M$ fields $\phi_1$, $M-1$ fields $\phi_2$, and a single, non-interacting third field $\phi_3$,
we find for the one-loop spectrum of Jordan blocks in the cyclic sector the (shifted) $q$-binomial coefficients
\begin{equation}\label{eq:Z3intro}
Z_{L,M}(q)
=\genfrac{[}{]}{0pt}{0}{L-1}{M-1}_q
=\prod_{k=1}^{M-1}\frac{q^{\frac{L-k}{2}}-q^{-\frac{L-k}{2}}}{q^{\frac{k}{2}}-q^{-\frac{k}{2}}}
\,.
\end{equation}
Interestingly, this result is also valid for the dynamical fishnet theory interaction Lagrangian \eqref{eq:Ldfn} (for generic couplings) due to the phenomenon of universality already pointed out in \cite{Ahn_2021}.

The paper is organised as follows. Section \ref{sec:modeldefinition} recalls the definitions of the non-hermitian chiral spin chain models at hand. The hypereclectic spin chain has a particularly simple Hamiltonian, essentially describing chiral right-movers on a chain along with a number of impenetrable non-movers, which we call \textit{walls}. Section \ref{sec:onewall} derives the exact solution of this model in the case of a single wall. The partition function method is introduced, and the solution is expressed in terms of $q$-binomial coefficients. Section \ref{sec:manywalls} generalises these findings to an arbitrary number of walls. Section \ref{sec:universality} analyses the generic three-parameter eclectic model, and, for the case of a single wall, sketches a proof of the universality hypothesis. Some remarks on the general case are made. We end with the short, concluding section \ref{sec:conclusions}, where it is also pointed out that the most important open issue seems to be our current inability to use integrability to analyse these integrable models. A few appendices \ref{app:shortening},\ref{app:universality},\ref{app:finetune} give further technical details.

\section{The (Hyper)eclectic Spin Chain}
\label{sec:modeldefinition}
In this section we collect basic facts about the models under consideration.
\subsection{Hamiltonian}
We consider local single-trace operators in the holomorphic 3-scalar sector of the theory \eqref{eq:Ldfn}
\begin{equation}
\mathcal{O}_{j_1,j_2,\dots,j_L}(x)=\text{tr}\left(\phi_{j_1}\phi_{j_2}\cdots\phi_{j_L}(x)\right), \hspace{1cm} j_i\in \{1,2,3\}.
\end{equation}
In $\mathcal{N}=4$ SYM the one-loop dilatation operator in the analogous sector can be written as a sum over permutation operators and enjoys an $\mathfrak{su}(3)$ symmetry \cite{Minahan_2003}. In the strongly twisted theory \eqref{eq:Ldfn} this symmetry is broken and the one-loop dilatation operator $H_{\text{ec}}: \left(\mathbb{C}^3\right)^{\otimes L}\rightarrow \left(\mathbb{C}^3\right)^{\otimes L}$ is a sum over \textit{chiral} permutation operators \cite{Ahn_2021}
\begin{equation}\label{eq:Hec}
H_{\text{ec}}=H_1+H_2+H_3=\sum_{i=1}^L \left(\xi_1\mathcal{H}_1^{i,i+1} + \xi_2\mathcal{H}_2^{i,i+1}+\xi_3\mathcal{H}_3^{i,i+1}\right) .
\end{equation}
The chiral permutation operators $\mathcal{H}_i:\mathbb{C}^3\otimes \mathbb{C}^3 \rightarrow \mathbb{C}^3\otimes \mathbb{C}^3$ act as follows:
\beq
\mathcal{H}_1\ket{32}=\ket{23},\qquad \mathcal{H}_2 \ket{13}=\ket{31}, \qquad \mathcal{H}_3 \ket{21}=\ket{12},
\eeq 
and annihilate all other states. Periodic boundary conditions are implemented $\mathcal{H}_i^{L,L+1}\equiv \mathcal{H}_i^{L,1}$. We have simplified the notation for the states of the spin chain by
\begin{equation}
\ket{\phi_{j_1}\phi_{j_2}\cdots \phi_{j_L}}\rightarrow \ket{j_1j_2\cdots j_L}.
\end{equation}
Therefore the Hamiltonian \eqref{eq:Hec} scans a state for neighboring fields in \textit{chiral order} $\ket{32}, \ket{13},$ or $\ket{21},$ and swaps them to \textit{anti-chiral order} $\ket{23}, \ket{31},$ and $\ket{12}$ respectively. E.g.\ we have
\begin{align}
&H_{\text{ec}}\ket{321}=\xi_1\ket{231}+\xi_3\ket{312}+\xi_2\ket{123},\\ \notag
&H_{\text{ec}}\ket{123}=0.
\end{align}
Setting $\xi_1=\xi_2=0, \xi_3\equiv \xi$ we recover the \textit{hypereclectic} model
\begin{equation}\label{eq:Hhec}
 H_3=\xi\sum_{i=1}^L\mathcal{H}_3^{i,i+1}.
\end{equation}
The Hamiltonians \eqref{eq:Hec} and \eqref{eq:Hhec} are block diagonal with respect to sectors of fixed numbers $K$ of $\phi_3$ fields, $M-K$ of $\phi_2$ fields, and $L-M$ of $\phi_1$ fields. We define $V^{L,M,K}$ to be the vector subspace of $ \left(\mathbb{C}^3\right)^{\otimes L}$ corresponding to these numbers of fields. Clearly we have
\begin{equation}
\text{dim} \hspace{0.1cm}V^{L,M,K}=\frac{L!}{(L-M)!(M-K)!K!}.
\end{equation}
$H_3$ corresponds to the one-loop dilatation operator in the fishnet theory, where we consider $K$ non-dynamical insertions $\phi_3$, which act as walls. For $K=0$ this operator, although non-Hermitian, is diagonalisable via a coordinate Bethe ansatz \cite{Ipsen_2019}. It corresponds essentially to a chiral version of the XY-model \cite{Lieb:1961fr}.
\subsection{Translation Operator and Cyclicity Classes}\label{sec:cyclicity}
We can further reduce the state space by considering the translation invariance of these Hamiltonians. Each $H_i$ commutes with the translation operator $U$
\begin{equation}
[H_i,U]=0, \qquad i=1,2,3,
\end{equation}
where $U$ generates a shift along the chain
\begin{equation}
U\ket{j_1j_2\cdots j_{L-1} j_L}=\ket{j_Lj_1j_2\cdots j_{L-1}}.
\end{equation}
This further implies $[H_{\text{ec}},U]=0$. Therefore we can choose to work in a basis where $U$ is diagonal. $U$ has $L$ distinct eigenvalues given by the $L^{th}$ roots of unity
\begin{equation}
\omega_L^k= e^{2\pi i k/L}, \hspace{1cm} k=0,1,\dots, L-1.
\end{equation} 
The $U$-eigenstates in $V^{L,M,K}$ with eigenvalue $\omega_L^k$ are said to be in the \textit{$k^{th}$ cyclicity class} $V^{L,M,K}_{k}$. The $k=0$ cyclicity class $V^{L,M,K}_{k=0}$ is known as the cyclic sector. The states in the $k^{th}$ cyclicity class are easily generated by acting repeatedly on a reference elementary state\footnote{We call single ket states $\ket{j_1j_2\dots j_L}$ elementary. In general states are linear combinations of these.} with $\omega_L^{-k} U$. For example, given $\ket{123}\in V^{3,2,1}$ we can form the cyclic state
\begin{equation}
\ket{123}+U\ket{123}+U^2\ket{123}=\ket{123}+\ket{312}+\ket{231},
\end{equation}
and states with $k=1$ or $k=2$
\begin{equation}
\ket{123}+\omega_3^{-1}U\ket{123}+\omega_3^{-2}U^2\ket{123}=\ket{123}+e^{-2\pi i/3}\ket{312}+e^{-4\pi i/3}\ket{231},
\end{equation}
\begin{equation}
\ket{123}+\omega_3^{-2}U\ket{123}+\omega_3^{-4}U^2\ket{123}=\ket{123}+e^{-4\pi i/3}\ket{312}+e^{-8\pi i/3}\ket{231}.
\end{equation}
For a given $L,M,K$ counting the number of states in $V^{L,M,K}$ with a given cyclicity $k$ requires P\'{o}lya counting, see for example \cite{2005sprol}. We denote the states in the $k^{th}$ cyclicity class by
\begin{equation}\label{eq:Ck}
\ket{j_1j_2\cdots j_L}_k \equiv \sum_{l=0}^{L-1}(w^{-k}U)^l \ket{j_1j_2\cdots j_L}\equiv \mathcal{C}_k \ket{j_1j_2\cdots j_L},
\end{equation}
where $\mathcal{C}_k$ is an (unnormalised) projector\footnote{Note that this projection may also result in a zero vector.} $\mathcal{C}_k^2\propto \mathcal{C}_k$ onto the $k^{th}$ cyclicity class $V^{L,M,K}_{k}$. For the hypereclectic spin chain we find it more natural to consider a so-called \textit{static} basis, which we describe at the beginning of section \ref{sec:onewall}.

\subsection{Spectral Problem}
Given a Hermitian Hamiltonian $H$ on an $n$-dimensional Hilbert space, it is well-known that one can construct an orthonormal $H$-eigenbasis $\psi_j$, $j=1,2,\dots, n$, such that
\begin{equation}
H\psi_j = E_j\psi_j \qquad j=1,2,\dots, n,
\end{equation}
where $E_j\in \mathbb{C}$ are the (possibly degenerate) eigenvalues of $H$.

For non-Hermitian Hamiltonians diagonalisability is not guaranteed, and indeed the (hyper)eclectic Hamiltonian is nilpotent and therefore non-diagonalisable in sectors with $K>0$. In this case there is still an essentially unique\footnote{Up to the ordering of the Jordan blocks.}  form to which the matrix can be brought, namely its \textit{Jordan normal form}. Furthermore, it is exactly the structure of the Jordan normal form which determines how the logarithms appear in the two-point functions \cite{Hogervorst_2017}. 

Let $H^{L,M,K}_{\text{ec}}$ be the eclectic Hamiltonian \eqref{eq:Hec} restricted to $V^{L,M,K}$. Then there exists a set of \textit{generalised eigenstates} $\psi^{m_j}_j$, $j=1,\dots,N$, $m_j=1,\dots, l_j$, which satisfy
\begin{equation}
H^{L,M,K}_{\text{ec}}\psi_j^k=\psi_j^{k-1}, \qquad H^{L,M,K}_{\text{ec}}\psi_j^1=0.
\end{equation}
We then say there are $N$ Jordan blocks labelled by $j$, each of length $l_j$. We call $\psi_j^{l_j}$ the \textit{top state} of the $j^{\text{th}}$ block. Each block has a true eigenstate $\psi^1_j$ of $H$ with eigenvalue $0$. In more general situations each block has a generalised eigenvalue $\mathcal{E}_j$ associated to it. However, in our case we have $\mathcal{E}_j=0$ for each $j$ 
since $H^{L,M,K}_{\text{ec}}$ is nilpotent. 
On a Jordan block of length $l$, $H^{L,M,K}_{\text{ec}}$ acts as the $l\times l$ matrix
\begin{equation}
J_l=\left(
\begin{array}{ccccc}
0 & 1 &    &            &  0 \\
    & 0 & 1 &            &      \\
    &    & 0 & \ddots &     \\
    &    &    & \ddots &  1 \\
 0 &    &    &            &  0
 \end{array}
\right).
\end{equation}
For the rest of this paper our goal will be to determine the Jordan block spectrum of $H_{\text{ec}}$ as a function of the sector labels $L,M,K$. This means finding the length and multiplicities of each of the blocks. For example, consider the sector $L=5, M=3, K=1$, which contains $30$ total states. For generic values\footnote{Interestingly, the couplings can be tuned to give a finer Jordan block decomposition, see appendix \ref{app:finetune}.} of the couplings $\xi_i$ there are $5$ Jordan blocks of length 5, and $5$ Jordan blocks of length 1. We denote this as
\begin{equation}
\text{JNF}_{531}=(5^5,1^5).
\end{equation}
\section{Hypereclectic with One Wall}\label{sec:onewall}
In this section we describe a method to determine the full Jordan block spectrum for the hypereclectic spin chain in sectors where $K=1$, i.e.\ there is a single, non-moving $\phi_3$ field, which acts as a fixed wall. In these sectors the model is equivalent to a chiral XY spin chain with open boundary conditions. The sizes and multiplicities of the Jordan blocks can be read off very simply from a generating function $Z_{L,M}(q)$, and the states of the Jordan blocks are determined by algorithmic methods. Throughout this section we denote the hypereclectic Hamiltonian $H_{3}\equiv H$ and set $\xi=1$. 

Since the $\phi_3$ field does not move under the action of $H$, we can further restrict to sectors with a fixed position of $\phi_3$. We will restrict to \textit{static} states of the form $\ket{j_1j_2\cdots j_{L-1}3}$, where $j_1,j_2,\dots, j_{L-1}\in\{1,2\}$. We will refer to the subspace of $V^{L,M,1}$ spanned by states of this form as $W^{L,M}$. We can access states where $\phi_3$ is in a different position by acting with the translation operator $U$, so that the Hilbert space decomposes
\begin{equation}\label{eq:VLM1}
V^{L,M,1}=\bigoplus_{j=0}^{L-1}U^j W^{L,M}.
\end{equation}
\subsection{Warmup Examples}
\paragraph{General $L$, $M=2$, $K=1$.}
The simplest situation is when $M=2$ and $K=1$. This means there are a single $\phi_3$ field, a single $\phi_2$ field, and $L-2$ $\phi_1$ fields. A natural basis for $W^{L,2}$ is given by $L-1$ states
\begin{equation}
\ket{211\cdots 113},\ket{121\cdots 113},\dots,\ket{111\cdots 123}.
\end{equation}
In this sector the states clearly form a single Jordan block of length $L-1$, as can be seen by acting with $H$ repeatedly on $\ket{211\cdots 113}$
\begin{equation}
 \ket{211\cdots 113}\xrightarrow{H}\ket{121\cdots 113} \xrightarrow{H}\cdots \xrightarrow{H}\ket{111\cdots 123}\xrightarrow{H} 0.
\end{equation}
We will refer to any state of the form $\ket{2^{M-K}1^{L-M}3^{K}}$ as \textit{anti-locked}, and  $\ket{1^{L-M}2^{M-K}3^{K}}$ as \textit{locked}. Similarly for the spaces $U^jW^{L,M}$, $j=1,\dots,L-1$ there is a single Jordan block of length $L-1$. Therefore for $M=2$ and $K=1$ we have
\begin{equation}
\text{JNF}_{L,2,1}=(L-1)^{L},
\end{equation}
 meaning there are $L$ blocks of length $L-1$.
\paragraph{$L=7, M=3, K=1$.}
The situation becomes more intricate with increasing $M$, which we illustrate with the example $L=7, M=3, K=1$. In this sector there are 4 $\phi_1$ fields, $2$ $\phi_2$ fields, and a single $\phi_3$ field. In $W^{7,3}$ there are 15 states. We use the important observation that the anti-locked state is always a top state for the longest Jordan block
 \begin{align}\label{eq:statetower}
     &\ket{2211113} &H^0\\
     \rightarrow &\ket{2121113}&H^1\notag\\
     \rightarrow &\ket{2112113}+\ket{1221113}&H^2\notag\\
     \rightarrow &\ket{2111213}+2\ket{1212113}&H^3\notag\\
     \rightarrow& \ket{2111123}+3\ket{1211213}+2\ket{1122113}&H^4\notag\\
     \rightarrow&4\ket{1211123}+5\ket{1121213}&H^5\notag\\
     \rightarrow& 5\ket{1112213}+9\ket{1121123}&H^6\notag\\
     \rightarrow &14\ket{1112123}&H^7\notag\\
     \rightarrow &14\ket{1111223}&H^8\notag\\
     \rightarrow&  0, &H^9 \notag
 \end{align}
so we have identified a Jordan block of length 9, whose eigenstate is proportional to the locked state $\ket{1111223}$. However since there are 15 states in the sector there must be additional Jordan blocks.

We note that each of the 15 elementary states appear in the tower of states \eqref{eq:statetower}. We classify these 15 states by where in this state tower they appear, by defining the \textit{level} $S$ of a state. We give the anti-locked state $\ket{2211113}$ $S=8$ and the locked state $\ket{1111223}$ $S=0$. In general, if an elementary state appears in the row $H^k$ of \eqref{eq:statetower}, we give it $S=8-k$. One notices that the $S$-value for a state is the total number of 1's to the right of each of the 2's. Defining $W_{S}^{7,3}$ to be the vector subspace of $W^{7,3}$ spanned by states with level $S$, we get
\begin{equation}
W^{7,3}=\bigoplus_{S=0}^8  W_{S}^{7,3},
\end{equation}
and it is clear that
\begin{equation}
H:W_{S}^{7,3}\rightarrow W_{S-1}^{7,3}, \qquad HW_{0}^{7,3}=0.
\end{equation}
In light of this, the next natural place to look for a top state of a Jordan block is in $W^{7,3}_{6}$. This is because a single state from each $W^{7,3}_{S}$ is already contained in the largest Jordan block, and $W^{7,3}_{6}$ is the space with largest $S$ with dimension larger than 1. We thus deduce that the top state for the next Jordan block must be of the form
\begin{equation}
\alpha\ket{2112113}+\beta\ket{1221113}\in W^{7,3}_6,
\end{equation}
where $\alpha\neq \beta$ as we want the state to be linearly independent from the corresponding state in the length 9 block. We act repeatedly on this state with $H$ until there is a possible choice for $\alpha$ and $\beta$ which makes the state vanish
    \begin{align*}
     &\alpha\ket{1221113}+\beta \ket{2112113}\\ &\rightarrow \beta\ket{2111213}+(\alpha+\beta)\ket{1212113}\\
     &\rightarrow \beta\ket{2111123}+(\alpha+2\beta)\ket{1211213}+(\alpha+\beta)\ket{1122113}\\
     &\rightarrow (\alpha+3\beta)\ket{1211123}+(2\alpha+3\beta)\ket{1121213}\\
     &\rightarrow (2\alpha+3\beta)\ket{1112213}+(3\alpha+6\beta)\ket{1121123}\\
     &\rightarrow (5\alpha+9\beta)\ket{1112123}.
       \end{align*}
We see that this yields a zero vector if $5\alpha+9\beta=0$, for example $\alpha=-9, \beta=5$. Therefore this chain of states determines a Jordan block of length 5, with top state $5\ket{2112113}-9\ket{1221113}\in W^{7,3}_6$ and eigenstate $-3\ket{1112213}+3\ket{1121123}\in W^{7,3}_{8-6}=W^{7,3}_2$.

There must be a single Jordan block of length 1 remaining, and by state counting this must be contained in $W_4^{7,3}$, since this is the only space with dimension greater than 2. We make the ansatz for the top state
\begin{equation}
\alpha' \ket{2111123}+\beta' \ket{1211213}+\gamma' \ket{1122113}\in W_4^{7,3}.
\end{equation}
This is easily checked to be an eigenstate of $H$ for $\alpha'=-\beta'=\gamma'=1$ and thus determines a Jordan block of length 1. The story is identical for the remaining spaces $U^jW^{7,3}$, $j=1,\dots,6$, so the overall Jordan normal form for $L=7, M=3, K=1$ is
\begin{equation}
\text{JNF}_{7,3,1}=(9^7,5^7,1^7).
\end{equation}
Let us step back and look at the state tower \eqref{eq:statetower}, from which we can see the dimensions
\begin{equation}
\text{dim}\hspace{0.05cm} W_S^{7,3},\hspace{1cm} S=0,1,\dots, 8
\end{equation}
by counting the number of elementary states in each row. We note that these dimensions form a diamond, in that they start from 1 at $S=8$, increase to a maximum of 3 at $S=4$, and decrease symmetrically to 1 at $S=0$. We encode these dimensions in a generating function
\begin{equation}\label{eq:Z73}
\bar{Z}_{7,3}(q)=\sum_{S=0}^{8}\text{dim}\hspace{0.05cm}W^{7,3}_S q^S=1+q+2q^2+2q^3+3q^4+2q^5+2q^6+q^7+q^8.
\end{equation}
Because of this diamond structure it is actually possible to deduce the Jordan block structure in $W^{7,3}$ from the generating function, a purely combinatorial object, up to some possible subtleties described in the next section. Given the generating function \eqref{eq:Z73} we identify the Jordan block of length 9 by the degree of the polynomial plus 1. We then subtract $1+q+q^2+\dots+q^8$ to represent the fact that there is one state at each level in this largest block. We then normalise the resulting polynomial to have lowest power $q^0$, and repeat the procedure:
\begin{align}\label{eq:procedure}
&1+q+2q^2+2q^3+3q^4+2q^5+2q^6+q^7+q^8\\ \rightarrow & 1+q+2q^2+q^3+q^4\notag\\ \rightarrow &1 \notag,
\end{align}
from which we deduce the Jordan block spectrum $(9,5,1)$. Therefore in the next section it will be our goal to generalise the arguments of this section and compute the generating function $\bar{Z}_{L,M}(q)$ for arbitrary $L,M$. 

\subsection{Generating Function}\label{generalLM}
For general $L,M$ we similarly grade the vector space in the static sector by the action of $H$
\begin{equation}\label{eq:directsum}
W^{L,M}=\bigoplus_{S=0}^{S_{\text{max}}}W_{S}^{L,M},
\end{equation}
\begin{equation}
H:W_{S}^{L,M}\rightarrow W_{S-1}^{L,M}, \qquad   H W_{0}^{L,M}=0.
\end{equation}
We have in general $S_{\text{max}}=L_1M_1$, where $L_1\equiv L-M$ is the number of 1's in the sector and $M_1\equiv M-1$ is the number of 2's. The anti-locked state is $\ket{2^{M_1}1^{L_1}3}\in W_{S_{\text{max}}}^{L,M}$ and the locked state is $\ket{1^{L_1}2^{M_1}3}\in W_{0}^{L,M}$. In general an elementary state takes the form
\beq\label{elembasis}
\ket{n_1,n_2,\dots,n_{M_1}}
\equiv\vert \underbrace{1\cdots 1}_{n_0}\mathbf{2}\underbrace{1\cdots 1}_{n_1}
\mathbf{2}\underbrace{1\cdots 1}_{n_2}\cdots\mathbf{2}\underbrace{1\cdots 1}_{n_{M_1}}\mathbf{3}\rangle,
\eeq
where $n_j$ is the number of $1$'s between the $j^{th}$ and $(j+1)^{th}$ 2. Clearly they should satisfy
\begin{equation}
\sum_{j=0}^{M_1}n_j=L-M=L_1.
\end{equation}
In this notation we can define the level $S$ for such a state which counts the number of 1's
on the right hand side of each of the 2's. Explicitly the state $
\ket{n_1,n_2,\dots,n_{M_1}}$ defined in \eqref{elembasis} has
\beq
\label{defs}
S=\sum_{j=1}^{M_1}jn_j.
\eeq
As before we define $W_{S}^{L,M}$ to be spanned by elementary states with this level $S$. The Hamiltonian acts on \eqref{elembasis} as 
\beq
\label{hamilhyper}
H: \ket{n_1,n_2,\dots,n_{M_1}}\quad\to\quad
\sum_{j=1}^{M_1}\ket{n_1,n_2,\dots,n_{j-1}+1,n_{j}-1,\dots,n_{M_1}}.
\eeq
\eqref{defs} and \eqref{hamilhyper} make it clear that $H$ decreases $S$ to $S-1$.

We now consider the problem of determining the dimensions of the spaces $W_S^{L,M}$. We would like to determine a generating function
\begin{equation}
\bar{Z}_{L,M}(q)=\sum_{S=0}^{S_{\text{max}}} \text{dim}\hspace{0.05cm} W^{L,M}_S q^S.
\end{equation}
These dimensions $\text{dim}\hspace{0.05cm} W^{L,M}_S$ are given by the number of partitions of the integer $S$ into at most $M_1$ parts, each less than or equal to $L_1$. Expressing \eqref{defs} as
\beq
\label{restricedpartition}
S=(n_1+n_2+\dots+n_{M_1})+ (n_2+\dots+n_{M_1})+\dots+n_{M_1}
\eeq
one can notice that there is one-to-one correpondence between an elementary vector in \eqref{elembasis} and such a restricted partition of $S$ in \eqref{restricedpartition}. For example, consider the case of the previous section, $L=7, M=3, K=1$. There were 3 elementary states in $W^{7,3}_4$:
\begin{align}
\ket{2111123}, \qquad (n_1+n_2, n_2)=(4,0),\\
\ket{1211213}, \qquad (n_1+n_2,n_2)=(3,1),\notag\\
\ket{1122113}, \qquad (n_1+n_2,n_2)=(2,2).\notag
\end{align}
These correspond to the partitions of the integer $4$ into at most $M_1=2$ parts, where each part is less than or equal to $L_1=4$. There are 3 such partitions $4=4=3+1=2+2$. 

Such restricted partitions described above can be generated by Gaussian (or $q$-) binomial coefficients \cite{stanley_2011} 
\beq
\bar{Z}_{L,M}(q)=\sum_{S=0}^{S_{\text{max}}} \text{dim}\hspace{0.05cm} W^{L,M}_S q^S=\binom{L-1}{M-1}_q=\prod_{k=1}^{M-1}\frac{1-q^{L-k}}{1-q^k},
\label{dim1}
\eeq
which is always a polynomial in $q$. Note that if we send $q\rightarrow 1$, the $q$-binomial reduces to the ordinary binomial coefficient and we have
\begin{equation}
\sum_{S=0}^{S_{\text{max}}} \text{dim}\hspace{0.05cm} W^{L,M}_S=\binom{L-1}{M-1}= \text{dim}\hspace{0.05cm} W^{L,M},
\end{equation}
as expected because of \eqref{eq:directsum}. \eqref{dim1} generates a list of dimensions\footnote{$\mathbf{d}_S=\mathbf{d}_S(L_1,M_1)$, we suppress the $L_1, M_1$ dependence for now.} $\mathbf{d}_S\equiv  \text{dim}\hspace{0.05cm} W^{L,M}_S$ 
\beq
\label{dimlist}
(\mathbf{d}_{S_{\rm max}},\mathbf{d}_{S_{\rm max}-1},\dots,\mathbf{d}_1,\mathbf{d}_0)\qquad {\rm with}\quad \mathbf{d}_0=\mathbf{d}_{S_{\rm max}}=1.
\eeq
Furthermore, from a property of the $q$-binomial coefficient, the dimensions are increasing from the left to the right until the midpoint, and decreasing after that, because of the symmetry
\beq
\mathbf{d}_S=\mathbf{d}_{\tilde S},\qquad {\tilde S}\equiv {S_{\rm max}-S}.
\eeq
For the space $W^{L,M}_{S_{\text{max}}}$, there is only one elementary state $\psi_0\equiv \ket{2^{M_1}1^{L_1}3}$, the anti-locked state.
By successive action of $H$,  a Jordan string of states is generated
\beq
\label{jordanstring1}
\psi_0\stackrel{H}{\longrightarrow}H\psi_0\stackrel{H}{\longrightarrow}H^2\psi_0\stackrel{H}{\longrightarrow}
\cdots\stackrel{H}{\longrightarrow}
H^{S_{\rm max}}\psi_0\stackrel{H}{\longrightarrow}0.
\eeq
Therefore, this generates a Jordan block of size $S_{\rm max}+1$, the largest one.

It turns out that the next dimension $\mathbf{d}_{S_{\rm max}-1}$ in \eqref{dimlist} is also one, as can be computed from \eqref{dim1}.
This means $W^{L,M}_{S_{\rm max}-1}$ is spanned by $H\psi_0$, the first descendant of the anti-locked state in \eqref{jordanstring1}. 
Therefore there is no other independent vector in $W^{L,M}_{S_{\rm max}-1}$ which can generate a new Jordan string. 

The top state of the second Jordan block arises at the first level $S=S_1$ below $S_{\text{max}}$ whose dimension is bigger than $1$. We can form $\mathbf{d}_{S_1}-1$ linearly independent potential top states in $W_{S_1}^{L,M}$, which are linearly independent from the $H$-descendant of the anti-locked state. We denote these states by $\psi^{(S_1)}_j$ ($j=1,\dots,\mathbf{d}_{S_1}-1$), and make the ansatz 
\beq
\psi^{(S_1)}_j=\sum_{i=1}^{\mathbf{d}_{S_1}}\alpha_j^{(i)}e_i^{(S_1)},
\eeq
where $e_i^{(S_1)}$ are the elementary states in $W^{L,M}_{S_1}$. $\alpha_j^{(i)}$ are constants which are determined by the condition that each $\psi^{(S_1)}_j$ constitutes a top state for a new Jordan block. Each of these states generates a Jordan string 
\beq
\psi^{(S_1)}_j\stackrel{H}{\longrightarrow}H\psi^{(S_1)}_j\stackrel{H}{\longrightarrow}H^2\psi^{(S_1)}_j\cdots\stackrel{H}{\longrightarrow}
H^{S_1-{\tilde S}_1}\psi^{(S_1)}_j\stackrel{H}{\longrightarrow}0,\quad j=1,\dots,\mathbf{d}_{S_1}-1.
\eeq
The condition $H^{S_1-{\tilde S}_1}\psi^{(S_1)}_j\stackrel{H}{\longrightarrow}0$ leads to a linear system of equations for the $\alpha_j^{(i)}$ which can be solved to determine the $\mathbf{d}_{S_1}-1$ new top states. These new Jordan blocks each have size $S_1-{\tilde S}_1+1$. The only possible subtlety is the potential for an `unexpected shortening' of the Jordan block, that is the possibility for the equation $H^{k}\psi^{(S_1)}_j=0$ to admit a solution in the $\alpha_j^{(i)}$ for some $k<S_1-\tilde{S}_1+1$. 
While we have not yet been able to rigorously disprove shortening in full generality, we have verified for a large number values of $L$ and $M$ that it does not happen. We were able to perform these extensive tests thanks to a mathematically more succinct reformulation of the problem, see appendix \ref{app:shortening} for details. We will assume that shortening cannot occur for the remainder of this paper.

The third set of Jordan blocks occurs at a level $S_2$, which is the largest integer satisfying $\mathbf{d}_{S_2}>\mathbf{d}_{S_1}$. 
Then, as before, we can form $\mathbf{d}_{S_2}-\mathbf{d}_{S_1}$ linearly independent potential top states which are linearly independent from $H$-descendants of the previous vectors, $\psi_0$ and $\psi^{(S_1)}_j$. We make a similar ansatz for these potential top states
\beq
\psi^{(S_2)}_j=\sum_{i=1}^{\mathbf{d}_{S_2}}\beta_j^{(i)}e_i^{(S_2)},
\eeq
 where $\beta_j^{(i)}$ are constants. These states create new
Jordan strings
\beq
\psi^{(S_2)}_j\stackrel{H}{\longrightarrow}H\psi^{(S_2)}_j\stackrel{H}{\longrightarrow}H^2\psi^{(S_2)}_j\cdots\stackrel{H}{\longrightarrow}
H^{S_2-{\tilde S}_2}\psi^{(S_2)}_j\stackrel{H}{\longrightarrow}0,\quad j=1,\dots,\mathbf{d}_{S_2}-\mathbf{d}_{S_1},
\eeq
and the final condition $H^{S_2-{\tilde S}_2}\psi^{(S_2)}_j\stackrel{H}{\longrightarrow}0$ is solved to determine the constants $\beta_{j}^{(i)}$. This leads to $\mathbf{d}_{S_2}-\mathbf{d}_{S_1}$ Jordan blocks of size $S_2-{\tilde S}_2+1$. This procedure can be continued until it reaches the maximum value of the dimension $\mathbf{d}_{S}$ which occurs at $S=[S_{\rm max}/2]$.

We note that for a given $L,M,$ the dimensions $\mathbf{d}_{S}$ are sufficient to determine the sizes and multiplicities of the Jordan blocks. For example, for $L=9, M=5$ we compute using \eqref{dim1}
   \begin{align}\label{eq:Z95}
\bar{Z}_{9,5}(q)=&1 + q + 2 q^2 + 3 q^3 + 5 q^4 + 5 q^5 + 7 q^6 + 7 q^7 + 8 q^8\\
   &+ 7 q^9 + 7 q^{10} + 5 q^{11} + 5 q^{12} + 3 q^{13} + 2 q^{14} + q^{15} + q^{16} \notag,
   \end{align} 
from which we can identify the Jordan normal form of $H$ in $W^{9,5}$ to be $(17,13,11,9^2,5^2,1)$ using the same procedure\footnote{With some practice one can easily and quickly `read off' the Jordan normal form from the generating function by visual inspection, i.e.\ this does not involve any calculations, just a bit of bookkeeping.} as \eqref{eq:procedure}. We can exhaust the Hilbert space by application of $U^j, j=1,\dots,8$, so that overall we have
\beq\text{JNF}_{9,5,1}=(17^9,13^9,11^9,9^{18},5^{18},1^9).
\eeq
For higher $K$, see the next section \ref{sec:manywalls}, we find it necessary to work with a slightly modified generating function for the dimensions of $W_{S}^{L,M}$ that is symmetric under $q\rightarrow q^{-1}$:
\beq \label{eq:zbar}
Z_{L,M}(q)=q^{-S_{\text{max}/2}}\bar{Z}_{L,M}(q)\equiv\genfrac{[}{]}{0pt}{0}{L-1}{M-1}_q.
\eeq
For example, \eqref{eq:Z73} and \eqref{eq:Z95} are modified to
\begin{equation}
Z_{7,3}(q)=q^{-4}+q^{-3}+2q^{-2}+2q^{-1}+3+2q+2q^2+q^3+q^4, 
\end{equation}
\begin{align}
Z_{9,5}(q)=q^{-8}&+q^{-7}+2q^{-6}+3q^{-5}+5q^{-4}+5q^{-3}+7q^{-2}+7q^{-1}+8 \\ 
  &+7q +7q^2+5q^3+5q^4+3q^5+2q^6+q^7+q^8. \notag
\end{align}
The modified function also provides an elegant way to determine the sizes and multiplicities of the Jordan blocks in a sector uniquely. We have
\begin{equation}\label{eq:extract}
Z_{L,M}(q) = \sum_{j} N_j [j]_q
=N_1\, q^0 + N_2\, \left( q^{-\frac{1}{2}}+q^{\frac{1}{2}} \right) 
+ N_3\, \left( q^{-1}+q^0+q^1 \right) +\ldots \,,
\end{equation}
where $N_j$ is the number of Jordan blocks of length $j$. $[j]_q$ is a modified $q$-number
\begin{equation}\label{eq:qnumber}
[j]_q = \frac{q^{j/2}-q^{-j/2}}{q^{1/2}-q^{-1/2}}= \sum_{k=\frac{-j+1}{2}}^{\frac{j-1}{2}} q^{k}.
\end{equation}
For example $Z_{7,3}(q)$ and $Z_{9,5}(q)$ can also be written
\begin{equation}
Z_{7,3}(q)=[1]_q+[5]_q+[9]_q,
\end{equation}
\begin{equation}
Z_{9,5}(q)=[1]_q+2[5]_q+2[9]_q+[11]_q+[13]_q+[17]_q,
\end{equation}
reflecting the Jordan block structures $(9,5,1)$ and $(17,13,11,9^2,5^2,1)$ respectively. A generating function which generates the Jordan block spectrum of all of $V^{L,M,1}$ can be obtained as a trace over the Hilbert space
\begin{equation}
\mathcal{Z}_{L,M}(q)=\text{tr}\hspace{0.05cm} q^{\hat{S}-S_{\text{max}}/2}=L \genfrac{[}{]}{0pt}{0}{L-1}{M-1}_q,
\end{equation}
where $\hat{S}$ acts on elementary states with well-defined values of $S$
\begin{equation}
\hat{S}\ket{S}=S\ket{S},
\end{equation}
and is extended by linearity.
\paragraph{Cyclicity classes.} We note that instead of considering states in $U^jW^{L,M}$ where the $\phi_3$ field is in a fixed position, we could have considered states in any cyclicity class $k$. If we replaced the states $\ket{j_1j_2\cdots j_{L-1}3}\rightarrow \mathcal{C}_k\ket{j_1j_2\cdots j_{L-1}3}$ for any $k=0,1,\dots,L-1$ the arguments of this section are unchanged because $[H,\mathcal{C}_k]=0$, where $\mathcal{C}_k$ is the unnormalised projector defined in \eqref{eq:Ck}. Therefore the Jordan normal form of $H$ is the same in $W^{L,M}$ and $V^{L,M,1}_k$ for any $k$.


\section{General Hypereclectic}
\label{sec:manywalls}
Here we discuss the extension of the previous section to sectors with many walls, i.e.\ $K>1$. The main observation is that $K>1$ states behave essentially like a tensor product of $K$ states with $K=1$. Any elementary state $v\in V^{L,M,K}$ ending in a 3 can be written
\begin{equation}\label{eq:v}
v=v_1\otimes v_2\otimes \cdots \otimes v_K,
\end{equation}
where $v_i\in W^{\ell_i+m_i+1,m_i+1}$ are elementary states themselves. We defined $W^{L,M}$ above \eqref{eq:VLM1}. $\ell_i$ denotes the number of 1's in $v_i$ and $m_i$ denotes the number of 2's. The hypereclectic Hamiltonian $H$ acts on states of the form \eqref{eq:v} as
\begin{equation}
Hv = Hv_1\otimes v_2 \otimes \cdots \otimes v_K+v_1\otimes Hv_2 \otimes \cdots \otimes v_K+\cdots+v_1\otimes v_2 \otimes \cdots \otimes Hv_K.
\end{equation}
We define $\boldsymbol{\ell}\equiv(\ell_1,\dots,\ell_K)$ and $\boldsymbol{m}\equiv(m_1,\dots,m_K)$, which should satisfy
\begin{equation}\label{eq:ellmcondition}
\sum_{i=1}^K \ell_i = L-M = L_1, \qquad \sum_{i=1}^K m_i = M-K = M_1.
\end{equation}
We will denote the spaces $\bigotimes_{i=1}^K W^{\ell_i+m_i+1,m_i+1}$ as \textit{subsectors}, and picking the vectors $\boldsymbol{\ell}, \boldsymbol{m}$ corresponds to a choice of subsector. We consider subsectors $(\boldsymbol{\ell}, \boldsymbol{m})$ satisfying \eqref{eq:ellmcondition} which are unique up to application of the translation operator $U^j$. In practise this means we identify $(\boldsymbol{\ell},\boldsymbol{m})\sim (\boldsymbol{\ell}',\boldsymbol{m}')$ if $\boldsymbol{\ell},\boldsymbol{\ell}'$ and $\boldsymbol{m},\boldsymbol{m}'$ are related by the same cyclic permutation $\sigma^n$
\beq
(\boldsymbol{\ell},\boldsymbol{m})\sim (\boldsymbol{\ell}',\boldsymbol{m}')\quad\longleftrightarrow\quad
(\boldsymbol{\ell}',\boldsymbol{m}') = (\sigma^n\boldsymbol{\ell},\sigma^n \boldsymbol{m}),
\eeq
\beq \label{eq:cyclic2}
\sigma(\ell_1,\ell_2,\dots,\ell_K)\equiv(\ell_2,\dots,\ell_K,\ell_1).
\eeq
In this way we can describe all the states in $V^{L,M,K}$ using the translation operator $U$. Overall we have
\begin{equation}\label{eq:VLMK}
V^{L,M,K}=\bigoplus_{(\boldsymbol{\ell},\boldsymbol{m})/\sim}\bigoplus_{j=1}^{L/S_{\boldsymbol{l},\boldsymbol{m}}} U^j \bigotimes_{i=1}^K W^{\ell_i+m_i+1,m_i+1},
\end{equation}
where we introduced the \textit{symmetry factor} for a subsector $S_{\boldsymbol{l},\boldsymbol{m}}$. The symmetry factor reflects the fact that some subsectors are especially symmetric with respect to cyclicity. This occurs when there is an $n<K$ such that
\beq
(\sigma^n \boldsymbol{\ell}, \sigma^n \boldsymbol{m})= (\boldsymbol{\ell}, \boldsymbol{m}),
\eeq
where $\sigma$ is the cyclic permutation defined in \eqref{eq:cyclic2}. In this case we give the subsector a symmetry factor $S_{\boldsymbol{\ell},\boldsymbol{m}}=K/n$. For example, let $L=14, M=8, K=4$ and take the subsector $\boldsymbol{\ell}=(2,1,2,1), \boldsymbol{m}=(1,1,1,1)$. We have $\sigma^2 \boldsymbol{\ell}=\boldsymbol{\ell}$ and $\sigma^2\boldsymbol{m}=\boldsymbol{m}$ and so $S_{\boldsymbol{\ell},\boldsymbol{m}}=4/2=2$ in this case. 

\subsection{Warmup Examples}
\paragraph{$L=7, M=4, K=2$.}
We begin with the simple example $L=7, M=4, K=2$. In this sector there are three $\phi_1$ fields, two $\phi_2$ fields, two $\phi_3$ fields and $\frac{7!}{3!2!2!}=210$ total states. In table \ref{table742} we show the 6 inequivalent choices of $(\boldsymbol{\ell},\boldsymbol{m})$, which corresponds to the 6 ways to decompose the states into $K=1$ states, on which $H$ acts block diagonally:
\begin{table}[h!]
\centering
\begin{tabular}{l|l|l|l|l}
                 & Form of state & Number of states &       $\quad\hspace{0.1cm}\boldsymbol{\ell},\boldsymbol{m} $       & JNF \\\hline
$\boldsymbol{1}$ & $\ket{111223}\otimes\ket{3}$                       & $10\times 1=10$              & $(3,0),(2,0)$                & $7\oplus 3$          \\
$\boldsymbol{2}$ & $\ket{11123}\otimes \ket{23}$                       & $4\times 1=4$                & $(3,0),(1,1)$              & 4                    \\
$\boldsymbol{3}$ & $\ket{11223}\otimes \ket{13}$                       & $6\times 1=6$              & $(2,1),(2,0)$           & $5\oplus 1$          \\
$\boldsymbol{4}$ & $\ket{1123}\otimes \ket{123}$                       & $3\times 2=6$      & $(2,1),(1,1)$  & $3\otimes 2$         \\
$\boldsymbol{5}$ & $\ket{1223}\otimes \ket{113}$                       & $3\times 1=3$               & $(1,2),(2,0)$                & 3                    \\
$\boldsymbol{6}$ & $\ket{1113}\otimes \ket{223}$                       & $1\times 1=1$                &  $(3,0),(0,2)$             & 1                   
\end{tabular}
\caption{\label{table742}Decomposition of $L=7, M=4, K=2$ states into $K=1$ states. The 3's should be regarded as fixed, whereas the 1's and 2's can be permuted within their ket.}
\end{table}
\newline
All subsectors except for $\boldsymbol{4}$ behave trivially as a single $K=1$ sector under the action of $H$. Their Jordan normal forms were determined in the previous section and are listed in the table. We look at states of the form $\boldsymbol{4}$ in a bit more detail. These states have the form of an $L=4, M=2, K=1$ state and an $L=3, M=2, K=1$ state glued together, which have Jordan blocks of size 3 and 2 respectively. The natural `anti-locked' state comes from gluing together the anti-locked states of the respective $K=1$ parts $\ket{2113213}$. We act successively on this state with $H$
\begin{align}
&\ket{2113213} \rightarrow \ket{1213213}+\ket{2113123}\\&\rightarrow \ket{1123213}+2\ket{1213123}\rightarrow 3\ket{1123123}\rightarrow 0 \notag,
\end{align}
which is a Jordan block of length 4. There is a further Jordan block of length 2 obtained by making the ansatz for a new top state
\begin{equation}
\gamma_1\ket{1213213}+\gamma_2\ket{2113123},
\end{equation}
and similarly to the last section this gives a Jordan block of length 2 for $\gamma_1=-1, \gamma_2=2$. Thus the Jordan decomposition of the subsector $\boldsymbol{4}$ is $(4,2)$. Since the Jordan decompositions of the $K=1$ sectors are $(3)$ and $(2)$ respectively, we denote this as $3\otimes 2=4\oplus 2$.

At the level of generating functions, we can deduce the Jordan normal form of the `tensor product' sectors by multiplying the generating functions of the corresponding $K=1$ sectors. For example, for the subsector $\boldsymbol{4}$ we have
\begin{align}
Z_{7,4,2}^{\boldsymbol{4}}(q)=Z_{4,2}(q)Z_{3,2}(q)=(q^{-1}+1+q)(q^{-1/2}+q^{1/2})\\=q^{-3/2}+2q^{-1/2}+2q^{1/2}+q^{3/2}, \notag
\end{align}
from which the Jordan normal form $(4,2)$ can be easily deduced using \eqref{eq:extract}. To obtain the full generating function for each of the subsectors in $L=7, M=4, K=2$ we can simply add the generating functions for each of the subsectors $\boldsymbol{1},\boldsymbol{2},\dots,\boldsymbol{6}$
\begin{align}
&Z_{7,4,2}(q)=\sum_{\boldsymbol{i}=\boldsymbol{1}}^{\boldsymbol{6}}Z_{7,4,2}^{\boldsymbol{i}}(q)\notag\\&=q^{-3}+2q^{-2}+2q^{-3/2}+4q^{-1}+3q^{-1/2}+6+3q^{1/2}+4q+2q^{3/2}+2q^2+q^3. \label{eq:Z742}
\end{align}
Using \eqref{eq:extract} leads to the following Jordan normal form:
\begin{equation}
\text{JNF}_{7,4,2}=(7,5,4^2,3^2,2,1^2).
\label{eq:JNFc742}
\end{equation}
In this sector there are no subtleties with cyclicity and the rest of the Hilbert space can be exhausted by application of the translation operator $U^j$, $j=1,\dots, 6$. For each $j$ we have the same argument as before, so the full Jordan block structure can be obtained as seven copies of \eqref{eq:JNFc742}
\begin{equation}
\text{JNF}^{\text{tot}}_{7,4,2}=(7^7,5^7,4^{14},3^{14},2^7,1^{14}).
\end{equation}
At the level of the generating function this can be obtained by multiplying \eqref{eq:Z742} by $L=7$. However, there are cases where cyclic symmetry leads to some subtleties, as we discuss next.
\paragraph{$L=8, M=4, K=2$.}
Let us consider the case of $L=8, M=4, K=2$. There are $\frac{8!}{4!4!2!}=420$ states in this sector. Therein one finds an ($\boldsymbol{\ell},\boldsymbol{m}$) subsector that is symmetric with respect to cyclicity. In table \ref{table842} we break the states into $K=1$ states as in the previous section,
\begin{table}[h!]
\centering
\begin{tabular}{l|l|l|l|l}
                 & Form of state & Number of states &$\quad\hspace{0.1cm}\boldsymbol{\ell},\boldsymbol{m}$    & Jordan decomposition \\\hline
$\boldsymbol{1}$ & $\ket{1111223}\otimes \ket{3}$                       & $15\times 1=15$               & $(4,0),(2,0)$                    & $9\oplus 5\oplus 1$          \\
$\boldsymbol{2}$ & $\ket{111223}\otimes\ket{13}$                       & $10\times 1=10   $             & $(3,1),(2,0)$             & $7\oplus 3$                   \\
$\boldsymbol{3}$ & $\ket{111123}\otimes\ket{23}$                       & $5 \times 1=5$              &  $(4,0),(1,1)$                 & $5$          \\
$\boldsymbol{4}$ & $\ket{11123}\otimes\ket{123}$                       & $4\times 2=8$      & $(3,1),(1,1)$   & $4\otimes 2=5\oplus 3$         \\
$\boldsymbol{5}$ & $\ket{11223}\otimes\ket{113}$                       & $6\times1=6$                & $(2,2),(2,0)$                 & $7\oplus 3$                    \\
$\boldsymbol{6}$ & $\ket{11113}\otimes\ket{223}$                       & $1\times 1=1$                &  $(4,0),(0,2)$            & 5                    \\
$\boldsymbol{7}$ & $\ket{1123}\otimes\ket{1123}$                       & $3\times 3=9$               & $(2,2),(1,1)$              & $3\otimes 3 = 5\oplus 3\oplus 1$          \\
$\boldsymbol{8}$ & $\ket{1223}\otimes\ket{1113}$                       & $3\times 1=3$      & $(1,3),(2,0)$  & $3$      
\end{tabular}
\caption{\label{table842}Decomposition of $L=8, M=4, K=2$ states into $K=1$ states.}
\end{table}
where we replaced $4\otimes 2=5\oplus 3$ and $3\otimes 3=5\oplus 3\oplus 1$ by multiplying the appropriate $K=1$ generating functions and naively extracting the resulting Jordan block structures using \eqref{eq:extract}. We see that $\boldsymbol{7}$ is the subsector where the issues with cyclicity emerge. For the other subsectors we can exhaust the rest of the state space by acting with $U^j, j=1,\dots,7$. However for subsector $\boldsymbol{7}$ applying the translation $U^4$ maps the states to a state in the same subsector, which reflects the fact this subsector has a symmetry factor $S_{\boldsymbol{\ell},\boldsymbol{m}}=2$. Therefore acting with $U^j, j=0,1,\dots,7$ leads to a double counting by a factor of 2. We can realise this at the level of an overall generating function for the $L=8, M=4, K=2$ sector by multiplying by $1/S_{\boldsymbol{\ell},\boldsymbol{m}}=1/2$ for the subsector $\boldsymbol{7}$
\begin{equation}
\mathcal{Z}_{8,4,2}(q)=8(Z_{8,4,2}^{\boldsymbol{1}}+Z_{8,4,2}^{\boldsymbol{2}}+Z_{8,4,2}^{\boldsymbol{3}}+Z_{8,4,2}^{\boldsymbol{4}}+Z_{8,4,2}^{\boldsymbol{5}}+Z_{8,4,2}^{\boldsymbol{6}}+\frac{1}{2}Z_{8,4,2}^{\boldsymbol{7}}+Z_{8,4,2}^{\boldsymbol{8}}).
\label{eq:Z842a}
\end{equation}
We compute \eqref{eq:Z842a} to be
\begin{equation}
\mathcal{Z}_{8,4,2}(q)=8q^{-4}+16q^{-3}+52q^{-2}+80q^{-1}+80q+52q^2+16q^3+8q^4.
\label{eq:Z842b}
\end{equation}
Using \eqref{eq:extract} we identify the Jordan normal form to be
\begin{equation}
\text{JNF}^{\text{tot}}_{8,4,2}=(9^8,7^8,5^{36},3^{28},1^{28}).
\end{equation}
\subsection{General \texorpdfstring{$L,M,K$}{L,M,K}}
 \label{sec:generalLMK}
Here we generalise the observations of the previous subsections to arbitrary $L,M,K$ sectors. Given an $L,M,K$ sector we consider a subsector $\bigotimes_{i=1}^K W^{\ell_i+m_i+1,m_i+1}$ defined by the vectors $\boldsymbol{\ell},\boldsymbol{m}$. The anti-locked state takes the form
\beq
\Omega=\ket{ (2\cdots 21\cdots 1)_1\mathbf{3}(2\cdots 21\cdots 1)_2\mathbf{3}\cdots (2\cdots 21\cdots 1)_K\mathbf{3}},
\eeq
where $(\ell_j,m_j)$ are the numbers of $1$'s and $2$'s in the $j^{\text{th}}$ bracket. Recall that we have
\beq
\quad \sum_{j=1}^K\ell_j=L_1=L-M,\quad 
\sum_{j=1}^Km_j=M_1=M-K.
\label{lmcondition}
\eeq
As for $K=1$, we can grade the vector space by the action of $H$
\beq
\bigotimes_{i=1}^K W^{\ell_i+m_i+1,m_i+1}=\bigoplus_{S=0}^{S_{\text{max}}}W^{\boldsymbol{\ell},\boldsymbol{m}}_S,
\eeq
where $W^{\boldsymbol{\ell},\boldsymbol{m}}_{S_{\text{max}}}$ is spanned by the anti-locked state and $H$ lowers the level $S\rightarrow S-1$. By acting successively with $H$ on $\Omega$, we will arrive at the locked state
\beq
\vert (1\cdots 12\cdots 2)_1\mathbf{3}(1\cdots 12\cdots 2)_2\mathbf{3}\cdots (1\cdots 12\cdots 2)_K\mathbf{3}\rangle.
\eeq
There will be many different configurations in the middle with lower values of $S$. For the anti-locked state we have 
\beq
S=S_{\rm max}=\boldsymbol{\ell}\cdot\boldsymbol{m}=\sum_{j=1}^K\ell_jm_j,
\eeq
and so the size of the largest Jordan block in each subsector is $S_{\rm max}+1$.
If we define the number of actions of $H$ on the $j^{\text{th}}$ bracket as $n_j$, a general state has a level
\beq
S=\sum_{j=1}^Ks_j=S_{\rm max}-N,\qquad s_j=\ell_jm_j-n_j, \quad N=\sum_{j=1}^Kn_j,\quad{\rm with}\quad 
0\le n_j\le \ell_jm_j.
\eeq
The anti-locked state has $S=S_{\rm max}$ (or $N=0$) and the locked state has $S=0$ (or $N=S_{\rm max}$).

\noindent Now consider states obtained by acting with $H$ $N$-times on the anti-locked state,
\beq
H^N\Omega=\sum_{n_1=0}^{\ell_1m_1}\cdots\sum_{n_K=0}^{\ell_Km_K}
\ket{H^{n_1}(2\cdots 21\cdots 1)\mathbf{3}H^{n_2}(2\cdots 21\cdots 1)\mathbf{3}\cdots H^{n_K}(2\cdots 21\cdots 1)\mathbf{3}}.
\eeq
The number of elementary states generated by each $H^{n_j}(2\cdots 21\cdots 1)$ was found in section \ref{generalLM} to be $\mathbf{d}_{l_jm_j-n_j}(\ell_j,m_j)$, which appeared as a coefficient of the $q$-binomial $\binom{\ell_j+m_j}{m_j}_q$ as defined in \eqref{dim1}. Therefore we can compute the number of elementary states at each level $S$ to be
\beq
\mathbf{D}^{\boldsymbol{\ell},\boldsymbol{m}}_{S_{\text{max}}-N}\equiv \text{dim}\hspace{0.05cm}W^{\boldsymbol{\ell},\boldsymbol{m}}_{S_{\text{max}}-N}=\sum_{n_1=0}^{\ell_1m_1}\cdots\sum_{n_K=0}^{\ell_Km_K}\prod_{j=1}^K\mathbf{d}_{l_jm_j-n_j}(\ell_j,m_j),
\qquad {\rm with}\quad \sum_{j=1}^Kn_j=N.\label{dimK}
\eeq
This can be recast into a generating function
\bea
\bar{Z}^{\boldsymbol{\ell},\boldsymbol{m}}(q)&=&\sum_{N=0}^{S_{\text{max}}}\mathbf{D}^{\boldsymbol{\ell},\boldsymbol{m}}_{S_{\text{max}}-N}\,q^N
=\sum_{N=0}^{S_{\text{max}}}\left[
\sum_{n_1=0}^{\ell_1m_1}\cdots\sum_{n_K=0}^{\ell_Km_K}\delta_{N,\sum_{i=1}^Kn_i}\prod_{j=1}^K\mathbf{d}_{l_jm_j-n_j}(\ell_j,m_j)
\right]\,q^{N}\notag\\
&=&
\sum_{n_1=0}^{\ell_1m_1}\cdots\sum_{n_K=0}^{\ell_Km_K}\prod_{j=1}^K\,\left[\mathbf{d}_{l_jm_j-n_j}(\ell_j,m_j)q^{n_j}\right]
=\prod_{j=1}^K\,
\left[\prod_{k=1}^{m_j}\frac{1-q^{\ell_j+m_j+1-k}}{1-q^k}\right], \label{gfunc}
\eea
using the expression for $K=1$ in \eqref{dim1}. This may be expressed through $q$-binomials as
\beq
\bar{Z}^{\boldsymbol{\ell},\boldsymbol{m}}(q)=\prod_{j=1}^K\,\binom{l_j+m_j}{m_j}_q.
\eeq
It proves that the generating function for an ${\boldsymbol{\ell}}, \boldsymbol{m}$ is simply a product of the corresponding $K=1$ generating functions. For example, if we take $L=13, M=7, K=3$ and consider the subsector $\boldsymbol{\ell}=(3,2,1), \boldsymbol{m}=(2,1,1)$ we find
\bea \label{eq:Z1373}
&\bar{Z}^{\boldsymbol{\ell},\boldsymbol{m}}(q)=\binom{5}{2}_q\binom{3}{1}_q\binom{2}{1}_q=\sum_{N=0}^{9}\mathbf{D}^{\boldsymbol{l},\boldsymbol{m}}_{9-N} q^N \\ \notag
&=1+3q+6q^2+9q^3+11q^4+11q^5+9q^6+6q^7+3q^8+q^9.
\eea
Analagously to the $K=1$ case, we can use \eqref{eq:procedure} to determine the Jordan block spectrum in this subsector
\begin{equation}
\text{JNF}^{\boldsymbol{\ell},\boldsymbol{m}}_{13,7,3}=(2^2,4^3,6^3,8^2,10).
\end{equation}
Since the states belonging to a given partition ${\boldsymbol{\ell}}, \boldsymbol{m}$ of $(L_1,M_1)$ are not mixed with those in a different partition, the total Jordan block spectrum is just direct sum of all the spectrum sets. 
One can sum over all inequivalent partitions formally. For this purpose, it is necessary to use the modified $q$-binomial coefficients defined in \eqref{eq:zbar}
\beq
Z^{\boldsymbol{\ell},\boldsymbol{m}}(q)=\prod_{j=1}^K \genfrac{[}{]}{0pt}{0}{\ell_j+m_j}{m_j}_q=\prod_{j=1}^K\,q^{-\ell_jm_j/2}\binom{l_j+m_j}{m_j}_q.
\eeq
For each $\boldsymbol{\ell},\boldsymbol{m}$ subsector we can exhaust the rest of the state space by acting with the translation operator $U^j, j=1,\dots, L-1$. The arguments of this section do not change in these cases, and so the overall generating function for a subsector can be obtained by simply multiplying it by $L$. The only exception is $\boldsymbol{\ell},\boldsymbol{m}$ subsectors which have a symmetry factor $S_{\boldsymbol{\ell},\boldsymbol{m}}\neq 1$. Adjusting for this possibility, we can define the generating function for a whole $L,M,K$ sector as a sum over inequivalent partitions
\beq\label{eq:ZLMKsum}
\mathcal{Z}_{L,M,K}(q)=\sum_{(\boldsymbol{\ell},\boldsymbol{m})/\sim}\frac{L}{S_{\boldsymbol{\ell},\boldsymbol{m}}}Z^{\boldsymbol{\ell},\boldsymbol{m}}(q).
\eeq
This total generating function gives the complete Jordan block spectrum, as in \eqref{eq:extract}:
\begin{equation}
\mathcal{Z}_{L,M,K}(q)= \sum_{j} N_j [j]_q
=N_1\, q^0 + N_2\, \left( q^{-\frac{1}{2}}+q^{\frac{1}{2}} \right) 
+ N_3\, \left( q^{-1}+q^0+q^1 \right) +\ldots \,.
\end{equation}
As for the $K=1$ case, $\mathcal{Z}_{L,M,K}(q)$ can alternatively be computed as a trace over the entire Hilbert space
\begin{equation}\label{eq:ZLMKtr}
\mathcal{Z}_{L,M,K}(q)=\text{tr} \hspace{0.05cm}q^{\hat{S}-\hat{S}_{\text{max}}/2},
\end{equation}
where $\hat{S}$ measures the level $S$ of an elementary state and $\hat{S}_{\text{max}}$ measures $S_{\text{max}}=\boldsymbol{\ell}\cdot\boldsymbol{m}$ of a state in an $\boldsymbol{\ell},\boldsymbol{m}$ subsector. Both operators are extended to the full Hilbert space by linearity. We can define $\hat{S}'\equiv \hat{S}-\hat{S}_{\text{max}}/2$ for brevity.
\paragraph{Cyclicity classes.}
The expression \eqref{eq:ZLMKtr}, which can also be expressed as \eqref{eq:ZLMKsum}, gives a generating function that describes the Jordan block spectrum of the hypereclectic model in an arbitrary sector of operators defined by $L, M, K$. However, in certain circumstances it might be useful to compute the Jordan block spectrum in a specific cyclicity class $k$, for example the cyclic sector $k=0$ relevant to quantum field theory. In this case, remarkably the formula \eqref{eq:ZLMKtr} still applies
\beq \label{eq:ZLMKtr2}
\mathcal{Z}^k_{L,M,K}(q)=\text{tr}_k \hspace{0.05cm}q^{\hat{S}'},
\eeq
where we take care to trace only over states of a fixed cyclicity $k$.

\section{Eclectic Spin Chain and Universality}
\label{sec:universality}
In the previous sections we described a method to find the full Jordan block spectrum of the hypereclectic model, as opposed to the eclectic model \eqref{eq:Hec} which is our main interest. However, we claim a \textit{universality hypothesis}: The Jordan block spectrum of the eclectic model for generic couplings $\xi_1,\xi_2,\xi_3$ is identical to that of the hypereclectic model, provided $L, M, K$ satisfy
\begin{equation}\label{eq:filling}
L_1=L-M\geq K, \qquad M_1=M-K\geq K.
\end{equation}
\eqref{eq:filling} implies that the number of $\phi_3$ fields in the sector does not exceed the number of $\phi_1$'s or $\phi_2$'s. Without loss of generality we can further take
\begin{equation}\label{eq:filling2}
L-M\geq M-K\geq K.
\end{equation}
Throughout this section we will consider \eqref{eq:filling2} to be satisfied, otherwise we can simply relabel the fields so that it is. It is possible to fine-tune the couplings to break down the Jordan block structure in certain cyclicity classes, as discussed in appendix \ref{app:finetune}. Since the $\phi_3$ fields no longer act as walls it is useful to work with states of a fixed cyclicity $k$, see section \ref{sec:cyclicity}. For definiteness in the following examples we will restrict to the cyclic sector $k=0$, which in addition happens to be the case relevant to quantum field theory.

\subsection{Eclectic Spin Chain and Level \texorpdfstring{$S$}{S}}
Recall that for elementary states in $K=1$ sectors we defined a level $S$, which corresponds to the total number of 1's to the right of each of the 2's in a state. Here we work with cyclic states
\begin{equation}\label{eq:cyclic}
\ket{j_1j_2\cdots j_{L-1}3}_{0}
\equiv\mathcal{C}_0\ket{j_1j_2\cdots j_{L-1}3}=\sum_{j=0}^{L-1}U^j \ket{j_1j_2\cdots j_{L-1}3}.
\end{equation}
We define $S$ in an analogous manner for states of the form \eqref{eq:cyclic}. For example the state $\ket{1211213}_0$ has $S=4$. Let us define $V_S$ to be the vector subspace of $V^{L,M,1}$ spanned by cyclic states with level $S$.\footnote{We suppress the $L,M$ dependence of $V_S$.} We saw previously that the hypereclectic Hamiltonian maps states in $V_S$ to $V_{S-1}$
\begin{equation}\label{eq:H3S}
H_3: V_S \rightarrow V_{S-1}, \qquad H_3 V_0=0.
\end{equation}
Let us investigate the action of the full eclectic Hamiltonian $H_{\text{ec}}= H_1 +  H_2 +  H_3$ on the vector spaces $V_S$. We find that
\begin{equation}\label{eq:H1H2S}
H_1: V_S\rightarrow V_{S-L_1}, \qquad H_2: V_{S}\rightarrow V_{S-M_1}.
\end{equation}
Since $L_1\geq M_1\geq 1$ \eqref{eq:H1H2S} implies that $H_1$ and $H_2$ decrease $S$ for a state by a greater than or equal amount to $H_3$. This already makes plausible that they will not affect the Jordan normal form of $H_3$, since $H_2$ and $H_1$ will annihilate states faster than $H_3$. For example, consider the anti-locked state $\ket{221113}_0\in V_6$ for $L=6, M=3, K=1$, so that $L_1=3, M_1=2$. Then
\begin{align}
H_1\ket{221113}_0=\ket{211123}_0\in V_3, \\
H_2\ket{221113}_0=\ket{122113}_0 \in V_4, \notag \\
H_3\ket{221113}_0=\ket{212113}_0 \in V_5. \notag
\end{align}
\subsection{Warmup Example}\label{sec:universalitywarmup}
Let us consider the eclectic model for $L=7, M=3, K=1$. In the hypereclectic model this sector has the Jordan block spectrum $(9,5,1)$ in $W^{9,5}$. Here we show that the eclectic model has the same Jordan block spectrum in the cyclic sector.
\paragraph{Top block.}
The anti-locked state in the cyclic sector $\ket{2211113}_0\in V_8$ again determines a Jordan block of length 9. The first descendant of the anti-locked state is
\begin{align}
H_{\text{ec}}\ket{2211113}_0=\xi_1 \ket{2111123}_0 + \xi_2 \ket{1221113}_0 + \xi_3 \ket{2121113}_0.
\end{align}
Note that the coefficients of $\xi_1, \xi_2,$ and $\xi_3$ are states with $S=4, 6,$ and $7$ respectively, which reflects equations \eqref{eq:H3S} and \eqref{eq:H1H2S}. In general acting with a power of $H_{\text{ec}}$ on $\ket{2211113}_0$ gives
\begin{equation}
H_{\text{ec}}^n \ket{2211113}_0 = H_3^n \ket{2211113}_0 \hspace{0.2cm}+ \hspace{0.2cm} \text{lower $S$ states}.
\end{equation}
It is then easy to see that $H^9\ket{2211113}_0=0$ and thus $\ket{2211113}_0$ is the top state for a Jordan block of length 9, as before.
\paragraph{Middle block.}
In the hypereclectic case the top state of the next Jordan block is
\begin{equation}\label{eq:top731}
\psi^{(6)}=-9\ket{1221113}_0+5\ket{2112113}_0 \in V_6,
\end{equation}
which satisfies $H_3^5\psi^{(6)}=0$. Thus $\psi^{(6)}$ determines a Jordan block of length 5 for $H_3$. However, in this case things are a bit trickier in the eclectic model. We have
\begin{equation}\label{eq:731res}
H_{\text{ec}}^5\psi^{(6)}=15\xi_2 \xi_3^4\ket{1111223}_0 \neq 0.
\end{equation}
It is however possible to modify the top state \eqref{eq:top731} by adding states of lower $S$, such that the residual term \eqref{eq:731res} vanishes. In this case it is sufficient to add states with $S=5$ to $\psi^{(6)}$. Since $\text{dim}\hspace{0.05cm}V_5=2$ we can add 2 states, to arrive at a new top state
\begin{equation}\label{eq:731modtop}
\chi^{(6)}= \psi^{(6)} + \gamma_1 \ket{1212113}_0 +\gamma_2 \ket{2111213}.
\end{equation}
This state satisfies
\begin{equation}
H_{\text{ec}}^5\psi^{(6)}=(-15\xi_2 +(5\gamma_1+4\gamma_2) \xi_3)\xi_3^4\ket{1111223}_0,
\end{equation}
which is 0 for $5\gamma_1+4\gamma_2=15\xi_2/\xi_3$. Note that this defines a one-parameter family of top states. Therefore the eclectic model also has a Jordan block of length 5 in this sector, with a slightly modified top state \eqref{eq:731modtop} which contains lower $S=5$ states.
\paragraph{Bottom block.}
In the hypereclectic model the top state for the final Jordan block is
\begin{equation}\label{eq:2top731}
\psi^{(4)}=\ket{2111123}_0- \ket{1211213}_0+ \ket{1122113}_0 \in V_4,
\end{equation}
which satisfies $H_3 \psi^{(4)}=0$ and thus determines a Jordan block of length 1. The action of the eclectic Hamiltonian on this state gives a residual
\begin{equation}\label{eq:731res2}
H_{\text{ec}}\psi^{(4)}=-\xi_1 \ket{1111223}_0-\xi_2\ket{1112213}_0-\xi_2\ket{1121123}_0\neq 0,
\end{equation} 
which consists of states with $S=0$ and $S=2$. As before we can eliminate this residual by adding states of lower $S$ to the top state \eqref{eq:2top731}. We first try to add states with $S=3$, and since $\text{dim}\hspace{0.05cm} V_3=2$ we add 2 states
\begin{equation}
\chi^{(4)}= \psi^{(4)} + \alpha_1 \ket{1121213}_0+\alpha_2\ket{1211123}_0.
\end{equation}
We check that for $\alpha_1=\xi_2/\xi_3, \alpha_2=-2\xi_2/\xi_3$ the $S=2$ states in the residual \eqref{eq:731res2} vanish
\begin{equation}
H_{\text{ec}}\chi^{(4)}=-\xi_1\ket{1111223}_0+\frac{\xi_2^2}{\xi_3}\ket{1112123}_0,
\end{equation}
which is a new residual consisting of an $S=1$ and an $S=0$ state. These can be removed by adding $S=2$ states into the top state ansatz
\begin{equation}
\bar{\chi}^{(4)}=\chi^{(4)}+\beta_1 \ket{1112213}_0+\beta_2 \ket{1121123}_0,
\end{equation}
and setting $\beta_1=\xi_1/\xi_2, \beta_2=-\xi_1/\xi_2-\xi_2^2/\xi_3^2$. With these choices for $\alpha_i$ and $\beta_i$ we have
\begin{equation}
H_{\text{ec}} \bar{\chi}^{(4)} = 0,
\end{equation}
and so we have identified the Jordan block of length one in this sector of the eclectic model. In summary, by taking a top state for the hypereclectic model at a level $S$, we can manufacture a top state (of a Jordan block of the same length) for the eclectic model by adding appropriate combinations of states with lower values of $S$. We will argue that it is always possible to add these states of lower $S$, thus rendering the Jordan block spectra of the hypereclectic and eclectic models equivalent.
\subsection{General Argument for \texorpdfstring{$K=1$}{K=1}}
Here we sketch a proof of universality for $K=1$, where the filling condition \eqref{eq:filling} is trivially satisfied, if all three particles are present in the spin chain state. We find it useful to first introduce the notion of \textit{supereclectic} models. These are intermediate models between the eclectic model $H_{\text{ec}}$ and the hypereclectic model $H_3$, defined by setting only a single coupling $\xi_1$ or $\xi_2$ equal to zero
\begin{equation}\label{eq:Hsuper}
H_{\text{super},i}= H_i + H_3, \qquad i=1,2.
\end{equation}
For both of these cases it is possible to prove rigorously that $H_{\text{super},i}$ has the same Jordan normal form as $H_3$ for generic couplings. The general strategy of the proof is reminiscent of the example given in section \ref{sec:universalitywarmup}. For the hypereclectic model, at a level $S$ satisfying $\mathbf{d}_{S}>\mathbf{d}_{S+1}$ we can construct $\mathbf{d}_{S}-\mathbf{d}_{S+1}$ top states 
\begin{equation}
\psi^{(S)}=\sum_{j=1}^{\mathbf{d}_S}\alpha_j^{(S)} e_j^{(S)},  
\end{equation}
where $\alpha_j^{(S)}$ are known coefficients and $e_j^{(S)}$ are the elementary states at level $S$. $\psi^{(S)}$ is the top state for a Jordan block of length $S-\tilde{S}+1$
\beq
H_3^{S-\tilde{S}+1}\psi^{(S)}=0,
\eeq
where we recall $\tilde{S}=S_{\text{max}}-S=(L-M)(M-1)-S$. We show that it is always possible to modify the state by adding a linear combination of states with lower $S$ 
\begin{equation}
\label{universalState}
\psi^{(S)}\to \psi_i^{(S)}=\psi^{(S)}+\sum_{n=0}^{S-1}\varphi^{(n)}
%
\end{equation}
where $\varphi^{(n)}\in V_n$. 
The modified state is a top state for a Jordan block of the same length in the supereclectic model $H_{\text{super},i}$
\begin{equation}\label{eq:Hsupernull}
H_{\text{super},i}^{S-\tilde{S}+1} \psi_i^{(S)}=0,
\end{equation}
which renders the Jordan normal forms of $H_{\text{super},i}$ and $H_3$ equivalent for generic couplings. 
More technical details of this proof are given in appendix \ref{app:universality}.
This argument can then be slightly modified to motivate that the Jordan normal forms of $H_{\text{ec}}$ and $H_3$ are also equivalent, see again \ref{app:universality}, even if we have not yet worked out all details of the proof.

\paragraph{Universality for \texorpdfstring{$K>1$}{K1}.}

It is even more complicated to show the universality for $K>1$. 
One main difference from the $K=1$ case of the supereclectic models, as 
explained in appendix \ref{app:universality}, is that the action of $h_j$ on $\varphi^{(S)}$ in general generates several states with differing $S$-values. 
If we interpret ${\hat S}(h_j\varphi^{(S)})$ in \eqref{Sofh} as the largest among these and replace $L_1$ with the
associated $\ell_j$, the same logic should be valid, so that one can construct for the supereclectic models
all subleading states in \eqref{dimK}.

For the eclectic model, however, a critical simplification used in \eqref{H1H2vanish} is not valid.
While we have extensive numerical evidence for general universality, we are currently unable to provide a proof. We leave this for future work.

\section{Conclusions and Outlook}
\label{sec:conclusions}

We introduced a generating function $\mathcal{Z}_{L,M,K}(q)$ that we conjectured (and partially proved) to fully enumerate the Jordan block spectrum of the hypereclectic model introduced in \cite{Ahn_2021}, for any sector of particles labelled by $L,M,K$. Interestingly, it takes a form reminiscent of a partition function, where one traces a certain kind of number operator over the state space.
It may also be expressed as a sum over products of (shifted) $q$-binomial coefficients, which elegantly reduces to a single $q$-binomial for the case of one wall, i.e.\ $K=1$. Furthermore, our approach for determining this generating function yields an algorithmic method for generating the states of the Jordan blocks. We also provided further strong evidence and partial proofs for the validity of the universality hypothesis of \cite{Ahn_2021}, i.e.\ the claim that the spectrum of the hypereclectic and eclectic models agree for special filling conditions. This is important, as the hypereclectic model is much easier to handle combinatorially in comparison with the eclectic one. Apart from its intrinsic value as a new type of solution for a new type of spin chain, our results appear to be an important starting point for an in-depth analysis of the indecomposability properties of the dynamical fishnet theory, cf.\ \eqref{eq:Ldfn}, an integrable logarithmic conformal field theory in four dimensions. In this context, note that $q$-binomials are ubiquitous in the analysis of two-dimensional logarithmic conformal field theories, see for example \cite{Flohr:2013dma}.

There are a number of gaps in our derivations that call for further research. Firstly, our combinatorial arguments do not rigorously exclude the possibility of `unexpected shortening' of Jordan blocks, as explained in \appref{app:shortening}. Secondly, while we made some progress towards a proof of the universality conjecture, a full proof is still missing. It is possible that the filling of these two gaps will require entirely new methods.

In this context, note that our results for these integrable models have {\it not} been derived by directly using integrability. Instead, they have been obtained by linear algebra arguments combined with combinatorics. 
Still, note that we were able to provide rather elegant formulas that clean up and organise to a large degree the (at first sight) incredibly intricate Jordan block structure of the models. One therefore wonders whether this, at least to us, rather astonishing fact is not an indirect manifestation of the integrability of these non-hermitian spin chain models. Understanding our findings from integrability is not only an interesting intellectual challenge, but might eventually allow to fill in the above mentioned gaps and incomplete proofs of this paper. Also, using integrability might lead to more explicit formulas than \eqref{eq:ZLMKsum} for $\mathcal{Z}_{L,M,K}(q)$ for $K>1$. 
 
 There are numerous further directions for investigations to consider. An interesting conceptual question is whether the Jordan block spectrum of other non-diagonalisable spin chains, integrable or not, can also be described by similar generating functions. Or else, is this something particular to the (hyper)eclectic spin chain? There would be a few natural ways to test this. For example, one could study the dilatation operator in other non-diagonalisable sectors of (dynamical) fishnet theory. These sectors could contain derivative fields/fermions, and would be more intricate to analyse. There are also different strong twist limits of $\mathcal{N}=4$ SYM available, which should contain new diagonalisable models, see \cite{Ahn_2021}. One could also consider the dilatation operator in the strong twist limit of ABJM theory \cite{caetano2018chiral}. In this case the first quantum correction to the dilatation operator appears at two loops, and we expect this would be a chiral version of the alternating spin chain given in \cite{2008min}.

The results of this paper concern the dilatation operator at one-loop order. It is natural to ask what might happen at higher loops. The dilatation operator certainly continues to be nilpotent, and therefore is non-diagonalisable. It would be interesting to see in detailed generality if and how the dilatation operator at different loop orders refines the Jordan block spectrum. And clearly if would be exciting to understand the structure of Jordan blocks on the non-perturbative level. Note that the QSC approach does not, in its current form, allow to even address the question \cite{2018QSC}.

\subsection*{Acknowledgements}
We are very thankful to Moritz Kade for helpful discussions, comments on the draft, and for writing a very useful Mathematica program for numerically obtaining the Jordan normal form of the hypereclectic spin chain.
We would like to express our sincere gratitude to the Brain Pool Program of the Korean National Research Foundation (NRF) under grant 2-2019-1283-001-1 for generous support of this research. MS thanks Ewha Womans University for hospitality in this difficult period. This project has received funding 
from NRF grant (NRF- 2016R1D1A1B02007258) (CA) and from the European Union's Horizon 2020 research and innovation programme under the Marie Sklodowska-Curie grant agreement  No.\ 764850 `SAGEX' (LC, MS). 

\appendix
\section{Unexpected Shortening}\label{app:shortening}
Here we succinctly reformulate the conditions for the unwanted `unexpected shortening' described in section \ref{generalLM}. This might be helpful for eventually finding a rigorous proof. In any case, it was very useful for the extensive numerical checking of our conjecture: unexpected shortening cannot happen.

In a sector with general $L,M$, $K=1$, we argued for the existence of a top state in $W^{L,M}_S$, where $S$ was such that $\mathbf{d}_{S}>\mathbf{d}_{S+1}:$
\begin{equation}
\psi^{(S)}=\sum_{i=1}^{\mathbf{d}_S}\alpha_i e_i^{(S)},
\end{equation}
where $\alpha_i$ are constants and $e_i^{(S)}$ are the elementary states in $W^{L,M}_S$. Acting with a power of $H$ on this state gives
\begin{equation}
H^k \psi^{(S)}= \sum_{i=1}^{\mathbf{d}_{S-k}}\sum_{j=1}^{\mathbf{d}_S}A^{(k)}_{ij}\alpha_j e_i^{(S-k)}=  \sum_{i=1}^{\mathbf{d}_{S-k}} (A^{(k)}\alpha)_i e_i^{(S-k)},
\end{equation}
where $A^{(k)}$ is a $\mathbf{d}_{S-k}\times\mathbf{d}_{S} $ matrix, and $\alpha$ is a vector of length $\mathbf{d}_S$ with entries $\alpha_i$. The top state $\psi^{(S)}$ defines a Jordan block of length $k$ if $H^{k}\psi^{(S)}=0$, or equivalently the homogeneous linear system
\begin{equation}\label{eq:homogeneous}
A^{(k)}\alpha =0
\end{equation}
admits at least one nontrivial solution in $\alpha$. We claim that the rank of $A^{(k)}$ is always maximal: 
\begin{equation}
\text{rank}(A^{(k)})=\text{min}(\mathbf{d}_{S-k},\mathbf{d}_{S} ).
\end{equation}
In this case, it is well-known that \eqref{eq:homogeneous} can only admit a nontrivial solution in $\alpha$ if and only if rank$(A^{(k)})<\mathbf{d}_{S}$. Moreover, the number of independent nontrivial solutions is $\mathbf{d}_{S}-\text{rank}(A^{(k)})$. Therefore a nontrivial solution only exists when $\mathbf{d}_{S-k}<\mathbf{d}_{S}$. This occurs precisely when $k=S-\tilde{S}+1$, as can be deduced from \eqref{dim1}. Therefore, if the rank of $A^{(k)}$ is always maximal, the top state $\psi^{(S)}$ determines $\mathbf{d}_{S}-\mathbf{d}_{S+1}$ Jordan blocks, each of length $S-\tilde{S}+1$. We checked the rank of $A^{(k)}$ for all top states and for all values of $k,$ up to $L=30, M=6,$ and always found it to be maximal, in line with our conjecture.




\section{Universality Details for \texorpdfstring{$K=1$}{K=1}}\label{app:universality}
In this section we prove that $H_{\text{super},i}$, defined in \eqref{eq:Hsuper}, has the same Jordan block structure as the hypereclectic model $H_3$ for $K=1$, under the assumption discussed in appendix \ref{app:shortening}. Then we describe how to modify these arguments to include the full eclectic Hamiltonian, and sketch a possible universality proof for $K=1$.
\paragraph{Universality for $H_{\text{super},1}$.}
We start with the first supereclectic model defined in \eqref{eq:Hsuper}, $H_{\text{super},1}$.
Consider a top vector $\psi^{(S)}$ for the hypereclectic model at a level $S$. This vector determines a Jordan block of length $n_S\equiv S-\tilde{S}+1$
\beq\label{eq:Hhectop}
H_3^{n_S}\psi^{(S)}=0.
\eeq
We can expand $H_{{\rm super},1}^{n_S}$ as
\beq
\label{Hexpansion}
H_{{\rm super},1}^{n_S}=\sum_{k=0}^{n_S}\binom{n_S}{k} H_1^kH_3^{n_S-k}=H_3^{n_S}+n_S H_1H_3^{n_S-1}
+\frac{n_S(n_S-1)}{2}H_1^2H_3^{n_S-2}+\cdots,
\eeq
where we have used $[H_1,H_3]=0$.
We introduce a shorthand notation 
\beq
\label{Hexpansion1}
H_{{\rm super},1}^{n_S}=\sum_{j=0}^{n_S}h_{j},\quad h_0=H_3^{n_S},\quad
h_{j}\equiv \binom{n_S}{k}  H_1^{j}H_3^{n_S-j},\quad j=1,\dots,n_S.
\eeq\t
Because of \eqref{eq:H3S} and \eqref{eq:H1H2S} each $h_j$ lowers the $S$-value of a state by $j(L_1-1)+n_S$. In other words, given a vector $\varphi^{(S)}\in V_S$ we have
\beq
\hat{S}(h_j \varphi^{(S)})=S-j(L_1-1)-n_S=(\tilde{S}-1)-j(L_1-1).
\label{Sofh}
\eeq
In particular, $h_j\varphi^{(S)}=0$ if this value is negative.
Now let us consider the $\tilde{S}$ value of a top vector to be in an interval
\beq
\ell(L_1-1)\le \tilde{S}-1<(\ell+1)(L_1-1).\label{interval1}
\eeq
In this case, all operators $h_j$ with $j>\ell$ will annihilate the top vector and its descendants.
Therefore, we may consider only operators $h_0,h_1,\dots,h_{\ell}$ and disregard others in \eqref{Hexpansion1}.

For this $S$ value of the top vector of the hypereclectic model $\psi^{(S)}$, 
we claim that we can construct a corresponding top vector $\psi^{(S)}_{1}$ of the supereclectic model $H_{\text{super},1}$, defined by
\beq \label{eq:supertop}
H_{{\rm super},1}^{n_S}\psi^{(S)}_{1}=0,
\eeq
via the ansatz
\beq
\psi^{(S)}_{1}=\varphi_0+\varphi_1+\cdots+\varphi_\ell,\qquad \varphi_0=\psi^{(S)},
\eeq
if the top vector has ${\tilde S}$ which satisfies \eqref{interval1}.
The condition \eqref{eq:supertop} can be written as
\bea
\label{topcondition}
&&(h_0+h_1+h_2+\cdots+h_{\ell})(\varphi_0+\varphi_1+\cdots+\varphi_\ell)\\ \notag
&=&(h_0\varphi_0)+(h_0\varphi_1+h_1\varphi_0)+\cdots+(h_0\varphi_\ell+h_1\varphi_{\ell-1}+\cdots+
h_{\ell}\varphi_0)+\cdots=0,
\eea
where we have grouped terms in a very particular way. The first term $h_0\varphi_0$ in \eqref{topcondition}  vanishes due to \eqref{eq:Hhectop}.
Now we want to find $\varphi_1$ in the second bracket from the restriction that it vanishes
\beq \label{consteq1}
h_0\varphi_1+h_1\varphi_0=0.
\eeq
Since $\hat{S}(h_1\varphi_0)=(\tilde{S}-1)-(L_1-1)$ from \eqref{Sofh}, this equation should be expressed 
by elementary vectors with this $S$ value. 
There are $\mathbf{d}_{(\tilde{S}-1)-(L_1-1)}$ of them, which becomes the number of constraints.\footnote{In fact, this is the maximum number of constraints since some of the elementary vectors may not appear.}
This equation also determines $\hat{S}(\varphi_1)=\hat{S}(h_1\varphi_0)+n_S=S-(L_1-1)$.
Therefore, $\varphi_1$ can be expressed as a linear combination of $\mathbf{d}_{S-(L_1-1)}$
elementary states. 
Since $\mathbf{d}_{S-(L_1-1)}=\mathbf{d}_{\tilde{S}+(L_1-1)}>\mathbf{d}_{(\tilde{S}-1)-(L_1-1)}$, one can 
solve coefficients of the linear combination from \eqref{consteq1} (not always unique).
This proves that we can always find the solution $\varphi_1$.

We require the next bracket in \eqref{topcondition} to vanish:
\beq
h_0\varphi_2+h_1\varphi_1+h_2\varphi_0=0.
\label{consteq2}
\eeq
Again, one can find that $\hat{S}(h_1\varphi_1)=\hat{S}(h_2\varphi_0)=(\tilde{S}-1)-2(L_1-1)$, from which we determine $\hat{S}(\varphi_2)=S-2(L_1-1)$.
Since the maximum number of constraints is smaller than that of the coefficients due to
$\mathbf{d}_{S-2(L_1-1)}>\mathbf{d}_{(\tilde{S}-1)-2(L_1-1)}$, one can find $\varphi_2$ from the
known vectors $\varphi_1$ and $\varphi_0$ using \eqref{consteq2}.

One can easily generalise this argument up to the $\ell$-th bracket in \eqref{topcondition}:
\beq
h_0\varphi_\ell+h_1\varphi_{\ell-1}+\cdots+h_{\ell}\varphi_0=0,
\label{consteq3}
\eeq
where the vectors $\varphi_{j},\ j=0,\dots,\ell-1$ have already been found in previous steps.
Since $\hat{S}(\varphi_j)=S-j(L_1-1)$ we have $\hat{S}(h_j\varphi_{\ell-j})=(\tilde{S}-1)-\ell(L_1-1)$ for $j=1,\dots,\ell$.
This determines  $S$-value of 
the unknown vector $\varphi_\ell$ to be $\hat{S}(\varphi_\ell)=S-\ell(L_1-1)$.
Again, the maximum number of constraints in \eqref{consteq3} is smaller than the number of coefficients in
the expansion of $\varphi_\ell$ in terms of elementary states, which guarantees that we can always find its solution.

There are more terms which we did not include in the second line of \eqref{topcondition}, but it is easy to show they all vanish.
For example, the $(\ell+1)$-th bracket would be
\beq
h_1\varphi_{\ell}+\cdots+h_{\ell}\varphi_1.
\eeq
Their $S$-values should be $(\tilde{S}-1)-(\ell+1)(L_1-1)$, which is negative due to \eqref{interval1}.
This means that all these vectors vanish.

This proves our universality conjecture for the  supereclectic model $H_{\text{super},1}$ by constructing the top vector explicitly as
\beq
\psi^{(S)}_{1}=\psi^{(S)}+\varphi_1+\cdots+\varphi_\ell,
\eeq
for $\tilde S$ in \eqref{interval1}.

Because $\tilde{S}\le S_{\rm max}/2$ ($\tilde{S}\le S$ by definition), the interval \eqref{interval1} is
limited by the maximum value of $\ell$ which is
\beq
\ell_{\rm max}=\left[\frac{L_1M_1}{2(L_1-1)}\right],
\eeq
where $[x]$ is the largest integer not exceeding $x$.

\paragraph{Universality for $H_{\text{super},2}$.}

The second supereclectic model $H_{\text{super},2}$ defined in \eqref{eq:Hsuper} can be analysed in exactly the same way.
Again, one can express 
\beq
\label{Hexpansion12}
H_{\text{super},2}^{n_S}=\sum_{m=0}^{n_S}g_{m},\quad
g_{m}\equiv\ \binom{n_S}{m} H_2^{m}H_3^{n_S-m},\quad m=0,\dots,n_S,\quad g_0=h_0=H_3^{n_S}.
\eeq
Each $g_m$ lowers $S$-values as follows:
\beq
\hat{S}(g_m \phi^{(S)})=S-m(M_1-1)-n_S.
\label{Sofh2}
\eeq
In the same way as before, a top vector with level $S$ (and corresponding $\tilde S$) with
\beq
m(M_1-1)\le \tilde{S}-1<(m+1)(M_1-1),\label{interval2}
\eeq
we only need to consider terms in \eqref{Hexpansion12} up to $g_{m}$.

The remaining procedure is identical to the previous case. 
One can always find ${\tilde\varphi}_k$ from ${\tilde\varphi}_0,\dots,{\tilde\varphi}_{k-1}$ using 
\beq
g_0{\tilde\varphi}_k+g_1{\tilde\varphi}_{k-1}+\dots+g_{k}{\tilde\varphi}_0=0,\quad k=1,\dots,m.
\label{consteq4}
\eeq
This proves the universality conjecture for $H_{\text{super},2}$ by constructing the top vector explicitly as
\beq
\psi^{(S)}_{2}=\psi^{(S)}+{\tilde\varphi}_1+\cdots+{\tilde\varphi}_m,
\eeq
for $\tilde S$ in \eqref{interval2}, where $m$ should be limited by the maximum value
\beq
m_{\rm max}=\left[\frac{L_1M_1}{2(M_1-1)}\right].
\eeq

\paragraph{Universality for General Eclectic Model.}

Powers of $H_{\rm ec}$ can be written as
\beq
\label{Hexpansioneclectic}
H_{\rm ec}^{n_S}=\sum_{k=0}^{n_S}\ \binom{n_S}{k}\, (H_1+H_2)^kH_3^{n_S-k}.
\eeq
This expression can be simplified greatly by observing that $H_1H_2=H_2H_1=0$ in sectors where $K=1$. 
This can be seen by acting with $H_1$ on any state
\bea
\label{H1H2vanish}
H_1\vert \mathbf{2}1\cdots 1
\mathbf{2}1\cdots 1\cdots\mathbf{2}1\cdots 1\mathbf{3}\rangle
=\vert1\cdots 1
\mathbf{2}1\cdots 1\cdots\mathbf{2}1\cdots 1\mathbf{2}\mathbf{3}\rangle.
\eea
Then, $H_2$ will annihilate the resulting state since it cannot contain $1\mathbf{3}$.
Therefore we can remove any terms with both $H_1$ and $H_2$
in the expansion \eqref{Hexpansioneclectic}, which leads to
\beq
\label{Hexpansioneclectic2}
H_{\rm ec}^{n_S}=h_0+(g_1+g_2+\cdots+g_{n_S})+(h_1+h_2+\cdots+h_{n_S}).
\eeq
We can restrict the interval for  $\tilde S$ by the two relations \eqref{interval1} and \eqref{interval2}.
Since $L_1\ge M_1$, for a given $\ell$ we can find $m$ such that
\beq
m(M_1-1)\le\ell(L_1-1)<(m+1)(M_1-1).
\eeq
In this case, the intersection of the two intervals is given by 
\beq
m(M_1-1)\le\ell(L_1-1)\le \tilde{S}-1<(m+1)(M_1-1).\label{interval3}
\eeq
For these values of $S$, the expansion of power of the eclectic Hamiltonian is truncated to
\beq
\label{Hexpansioneclectic3}
H_{\rm ec}^{n_S}=h_0+(g_1+g_2+\cdots+g_m)+(h_1+h_2+\cdots+h_\ell).
\eeq
We now claim that the top vector of the eclectic model can be always constructed from the hypereclectic top state $\psi^{(S)}=\varphi_0$ as follows:
\beq
\psi^{(S)}_{\rm ec}=\varphi_0+\sum_{i=1}^{m}{\tilde\varphi}_i+\sum_{j=1}^{\ell}\varphi_j.
\label{eclectictopvector}
\eeq
Let us provide the detailed proof for the simplest case $m=2,\ell=1$, with
\beq
2(M_1-1)\le(L_1-1)\le \tilde{S}-1<3(M_1-1).\label{interval4}
\eeq
We will show that the top vector for the eclectic model can be constructed as
\beq
\psi^{(S)}_{\rm ec}=\varphi_0+{\tilde\varphi}_1+{\tilde\varphi}_2+\varphi_1.
\eeq
One can expand $H_{\rm ec}^{n_S}\psi_{\text{ec}}^{(S)}=0$ as
\bea
&&(h_0+g_1+g_2+h_1)(\varphi_0+{\tilde\varphi}_1+{\tilde\varphi}_2+\varphi_1)
=(h_0\varphi_0)+(g_0{\tilde\varphi}_1+g_1\varphi_0)+\nonumber\\
&+&(g_0{\tilde\varphi}_2+g_1{\tilde\varphi}_1+g_2\varphi_0)
+(g_0\varphi_1+g_1{\tilde\varphi}_2+g_2{\tilde\varphi}_1+h_1\varphi_0)+\cdots=0.
\label{eclecticexpand}
\eea
The first three brackets in \eqref{eclecticexpand} have already been solved for $H_{\text{super},2}$,
therefore we only need to consider the fourth term and ellipsis.
The $S$-values of each term have already been computed as
\beq
\hat{S}(g_1{\tilde\varphi}_2)=\hat{S}(g_2{\tilde\varphi}_1)=({\tilde S}-1)-3(M_1-1)<
\hat{S}(h_1\varphi_0)=({\tilde S}-1)-(L_1-1).
\eeq
Therefore, $\varphi_1$ can be determined from $\varphi_0$ in the same way as for $H_{\text{super},1}$ with
additional subleading terms in $S$ from the known ${\tilde\varphi}_1,{\tilde\varphi}_2$.
The terms in the ellipsis in \eqref{eclecticexpand} are
\beq
\cdots=g_1\varphi_1+g_2{\tilde\varphi}_2+h_1{\tilde\varphi}_1+g_2{\varphi}_1+h_1{\tilde\varphi}_2+
h_1{\varphi}_1.
\label{ellipsis2}
\eeq
The $S$-values for these vectors are given by
\bea
&&\hat{S}(g_i\varphi_j)=\hat{S}(h_j{\tilde\varphi}_i)=({\tilde S}-1)-j(L_1-1)-i(M_1-1),\nonumber\\
&&\hat{S}(h_i\varphi_j)=({\tilde S}-1)-(i+j)(L_1-1),\quad
\hat{S}(g_i{\tilde\varphi}_j)=({\tilde S}-1)-(i+j)(M_1-1).
\eea
It is not difficult to see from \eqref{interval4} that all these vectors should vanish since their $S$-values are all negative. 

This procedure can now be generalised in principle to any value of $(\ell,m)$, although it is hard to give general, explicit expressions, since the mixed interval depends closely on explicit vaues of $L_1,M_1$. It would be interesting to complete the details of this sketch of a proof of $K=1$ universality.

\section{Fine Tuning and Cyclicity Classes}\label{app:finetune}
Although we have proven the universality hypothesis for generic values of the couplings $\xi_i$ for $K=1$, it is possible to fine-tune the couplings to destroy the Jordan block structures in a particular cyclicity class. We give a simple example of this occurring, for the sector $L=5, M=3, K=1$. There are 30 states in this sector:
\begin{align}
\mathcal{C}_k \ket{22113}, \quad  \mathcal{C}_k \ket{21213}, \quad \mathcal{C}_k \ket{12213}, \\
\mathcal{C}_k \ket{21123}, \quad  \mathcal{C}_k \ket{12123}, \quad \mathcal{C}_k \ket{11223}, \notag
\end{align}
where $\mathcal{C}_k$ is the unnormalised projector defined in \eqref{eq:Ck} and $k=0,1,2,3,4$ labels the cyclicity class. In each cyclicity class $k$ the hypereclectic model $H_3$ has Jordan decomposition $(5,1)$, so that the overall Jordan decomposition is $(5^5,1^5)$. The other models related to $H_3$ by permutations of the fields $H_1$ and $H_2$ have Jordan decomposition (3,2,1) in each cyclicity class. For generic $\xi_i$ we have argued that the eclectic Hamiltonian $H_{\text{ec}}=H_1 +  H_2 + H_3$ also has the Jordan decomposition $(5^5, 1^5)$, since this sector satisfies the filling conditions \eqref{eq:filling2}. Setting $\xi_3=0$ leads to a Jordan decomposition $(3^5,2^5,1^5)$. Interestingly, this decomposition can be further refined by tuning $\xi_1$ and $\xi_2$. Let us act with $H_{\text{ec}}|_{\xi_3=0}$ on the top state $\mathcal{C}_k\ket{22113}$:

\begin{align}
\mathcal{C}_k\ket{22113}&\rightarrow \omega^{k} \xi_1\mathcal{C}_k\ket{21123} + \omega^{-k}\xi_2 \mathcal{C}_k\ket{12213} \\
&\rightarrow (\omega^{2k}\xi_1^2+\omega^{-2k}\xi_2^2 ) \mathcal{C}_k \ket{11223}\rightarrow 0, \notag
\end{align}
where $\omega=e^{2\pi i/5}$ and we used $  \mathcal{C}_k U^{\pm 1}\psi=\omega^{\pm k} \mathcal{C}_k \psi, [H_i, \mathcal{C}_k]= 0$. For generic couplings this gives a length 3 block in each cyclicity class. However, if we tune the couplings such that $\xi_2^2 = -\omega^{4k}\xi_1^2$ the block splits into a 2-block and a 1-block in this cyclicity class $k$. There are two further top states in this sector:
\begin{align}
&\mathcal{C}_k\ket{21213} \rightarrow (\xi_1 \omega^{k}+\xi_2 \omega^{-k})\mathcal{C}_k \ket{12123}\rightarrow 0, \\
&\xi_2 \omega^{-k}\mathcal{C}_k \ket{21123}-\xi_1 \omega^{k}\mathcal{C}_k \ket{12213}\rightarrow 0. \notag
\end{align}
The first of these is a 2-block, which can be broken into two 1-blocks in a single cyclicity class if 
$\xi_2 = -  \omega^{2 k} \xi_1$.
The next of these is always a 1-block. From this example we see explicitly that finer Jordan block decompositions can be obtained in specific cyclicity classes by tuning the couplings appropriately.

\pdfbookmark[1]{\refname}{references}
\bibliographystyle{nb}
\bibliography{EclecticPaper}

\begin{thebibliography}{10}
\providecommand{\href}[2]{#2}
\providecommand{\arxivref}[2]{\href{http://arxiv.org/abs/#1}{#2}}
\providecommand{\doiref}[2]{\href{http://dx.doi.org/#1}{#2}}
\providecommand{\nbbstauthor}[1]{#1}
\providecommand{\nbbstjournal}[1]{\textsf{#1}}
\providecommand{\nbbsttitle}[1]{\textit{#1}}
\providecommand{\nbbsturl}[1]{\texttt{#1}}
\providecommand{\nbbsteprint}[1]{\texttt{#1}}
\providecommand{\nbbststyle}{\raggedright\small\parskip0pt}
\nbbststyle

\bibitem{Ipsen_2019}
\nbbstauthor{A.~C.~Ipsen, M.~Staudacher and L.~Zippelius},
\nbbsttitle{``The one-loop spectral problem of strongly twisted $\mathcal{N}$ =
  4 Super Yang-Mills theory''},
\nbbstjournal{\doiref{10.1007/jhep04(2019)044}{Journal~of~High~Energy~Physics~2019,~L.~Zippelius~(2019)}},
\nbbsteprint{\arxivref{1812.08794}{arxiv:1812.08794}},
\href{http://dx.doi.org/10.1007/JHEP04(2019)044}{\nbbsturl{http://dx.doi.org/10.1007/JHEP04(2019)044}}.

\bibitem{Ahn_2021}
\nbbstauthor{C.~Ahn and M.~Staudacher},
\nbbsttitle{``The integrable (hyper)eclectic spin chain''},
\nbbstjournal{\doiref{10.1007/jhep02(2021)019}{Journal~of~High~Energy~Physics~2021,~M.~Staudacher~(2021)}},
\nbbsteprint{\arxivref{2010.14515}{arxiv:2010.14515}},
\href{http://dx.doi.org/10.1007/JHEP02(2021)019}{\nbbsturl{http://dx.doi.org/10.1007/JHEP02(2021)019}}.

\bibitem{Frolov_2005}
\nbbstauthor{S.~Frolov},
\nbbsttitle{``Lax pair for strings in Lunin-Maldacena background''},
\nbbstjournal{\doiref{10.1088/1126-6708/2005/05/069}{Journal~of~High~Energy~Physics~2005,~069~(2005)}},
\nbbsteprint{\arxivref{hep-th/0503201}{hep-th/0503201}},
\href{http://dx.doi.org/10.1088/1126-6708/2005/05/069}{\nbbsturl{http://dx.doi.org/10.1088/1126-6708/2005/05/069}}.

\bibitem{Frolov2_2005}
\nbbstauthor{S.~Frolov, R.~Roiban and A.~Tseytlin},
\nbbsttitle{``Gauge-string duality for (non)supersymmetric deformations of
  super-Yang-Mills theory''},
\nbbstjournal{\doiref{10.1016/j.nuclphysb.2005.10.004}{Nuclear~Physics~B~731,~1~(2005)}},
\nbbsteprint{\arxivref{hep-th/0507021}{hep-th/0507021}},
\href{http://dx.doi.org/10.1016/j.nuclphysb.2005.10.004}{\nbbsturl{http://dx.doi.org/10.1016/j.nuclphysb.2005.10.004}}.

\bibitem{Gurdogan_2016}
\nbbstauthor{O.~Gurdogan and V.~Kazakov},
\nbbsttitle{``New Integrable 4D Quantum Field Theories from Strongly Deformed
  Planar N=4 Supersymmetric Yang-Mills Theory''},
\nbbstjournal{\doiref{10.1103/physrevlett.117.201602}{Physical~Review~Letters~117,~V.~Kazakov~(2016)}},
\nbbsteprint{\arxivref{1512.06704}{arxiv:1512.06704}},
\href{http://dx.doi.org/10.1103/PhysRevLett.117.201602}{\nbbsturl{http://dx.doi.org/10.1103/PhysRevLett.117.201602}}.

\bibitem{Kazakov_2019}
\nbbstauthor{V.~Kazakov, E.~Olivucci and M.~Preti},
\nbbsttitle{``Generalized fishnets and exact four-point correlators in chiral
  CFT4''},
\nbbstjournal{\doiref{10.1007/jhep06(2019)078}{Journal~of~High~Energy~Physics~2019,~M.~Preti~(2019)}},
\nbbsteprint{\arxivref{1901.00011}{arxiv:1901.00011}},
\href{http://dx.doi.org/10.1007/JHEP06(2019)078}{\nbbsturl{http://dx.doi.org/10.1007/JHEP06(2019)078}}.

\bibitem{Gromov_2018}
\nbbstauthor{N.~Gromov, V.~Kazakov, G.~Korchemsky, S.~Negro and G.~Sizov},
\nbbsttitle{``Integrability of conformal fishnet theory''},
\nbbstjournal{\doiref{10.1007/jhep01(2018)095}{Journal~of~High~Energy~Physics~2018,~G.~Sizov~(2018)}},
\nbbsteprint{\arxivref{1706.04167}{arxiv:1706.04167}},
\href{http://dx.doi.org/10.1007/JHEP01(2018)095}{\nbbsturl{http://dx.doi.org/10.1007/JHEP01(2018)095}}.

\bibitem{Gromov3_2019}
\nbbstauthor{N.~Gromov, V.~Kazakov and G.~Korchemsky},
\nbbsttitle{``Exact correlation functions in conformal fishnet theory''},
\nbbstjournal{\doiref{10.1007/jhep08(2019)123}{Journal~of~High~Energy~Physics~2019,~G.~Korchemsky~(2019)}},
\nbbsteprint{\arxivref{1808.02688}{arxiv:1808.02688}},
\href{http://dx.doi.org/10.1007/JHEP08(2019)123}{\nbbsturl{http://dx.doi.org/10.1007/JHEP08(2019)123}}.

\bibitem{Chicherin_2017}
\nbbstauthor{D.~Chicherin, V.~Kazakov, F.~Loebbert, D.~M{\"u}ller and
  D.-l.~Zhong},
\nbbsttitle{``Yangian symmetry for fishnet Feynman graphs''},
\nbbstjournal{\doiref{10.1103/physrevd.96.121901}{Physical~Review~D~96,~D.~(2017)}},
\nbbsteprint{\arxivref{1708.00007}{arxiv:1708.00007}},
\href{http://dx.doi.org/10.1103/PhysRevD.96.121901}{\nbbsturl{http://dx.doi.org/10.1103/PhysRevD.96.121901}}.

\bibitem{Loebbert_2020}
\nbbstauthor{F.~Loebbert, D.~M{\"u}ller and H.~M{\"u}nkler},
\nbbsttitle{``Yangian bootstrap for conformal Feynman integrals''},
\nbbstjournal{\doiref{10.1103/physrevd.101.066006}{Physical~Review~D~101,~H.~M{\"u}nkler~(2020)}},
\nbbsteprint{\arxivref{1912.05561}{arxiv:1912.05561}},
\href{http://dx.doi.org/10.1103/PhysRevD.101.066006}{\nbbsturl{http://dx.doi.org/10.1103/PhysRevD.101.066006}}.

\bibitem{Corcoran_2021}
\nbbstauthor{L.~Corcoran, F.~Loebbert, J.~Miczajka and M.~Staudacher},
\nbbsttitle{``Minkowski box from Yangian bootstrap''},
\nbbstjournal{\doiref{10.1007/jhep04(2021)160}{Journal~of~High~Energy~Physics~2021,~M.~Staudacher~(2021)}},
\nbbsteprint{\arxivref{2012.07852}{arxiv:2012.07852}},
\href{http://dx.doi.org/10.1007/JHEP04(2021)160}{\nbbsturl{http://dx.doi.org/10.1007/JHEP04(2021)160}}.

\bibitem{Basso_2017}
\nbbstauthor{B.~Basso and L.~J.~Dixon},
\nbbsttitle{``Gluing Ladder Feynman Diagrams into Fishnets''},
\nbbstjournal{\doiref{10.1103/physrevlett.119.071601}{Physical~Review~Letters~119,~L.~J.~Dixon~(2017)}},
\nbbsteprint{\arxivref{1705.03545}{arxiv:1705.03545}},
\href{http://dx.doi.org/10.1103/PhysRevLett.119.071601}{\nbbsturl{http://dx.doi.org/10.1103/PhysRevLett.119.071601}}.

\bibitem{Derkachov_2020}
\nbbstauthor{S.~Derkachov and E.~Olivucci},
\nbbsttitle{``Exactly Solvable Magnet of Conformal Spins in Four Dimensions''},
\nbbstjournal{\doiref{10.1103/physrevlett.125.031603}{Physical~Review~Letters~125,~E.~Olivucci~(2020)}},
\nbbsteprint{\arxivref{1912.07588}{arxiv:1912.07588}},
\href{http://dx.doi.org/10.1103/PhysRevLett.125.031603}{\nbbsturl{http://dx.doi.org/10.1103/PhysRevLett.125.031603}}.

\bibitem{Basso_2021}
\nbbstauthor{B.~Basso, L.~J.~Dixon, D.~A.~Kosower, A.~Krajenbrink and
  D.-l.~Zhong},
\nbbsttitle{``Fishnet four-point integrals: integrable representations and
  thermodynamic limits''},
\nbbstjournal{\doiref{10.1007/jhep07(2021)168}{Journal~of~High~Energy~Physics~2021,~D.~(2021)}},
\nbbsteprint{\arxivref{2105.10514}{arxiv:2105.10514}},
\href{http://dx.doi.org/10.1007/JHEP07(2021)168}{\nbbsturl{http://dx.doi.org/10.1007/JHEP07(2021)168}}.

\bibitem{Gromov_2019}
\nbbstauthor{N.~Gromov and A.~Sever},
\nbbsttitle{``Derivation of the Holographic Dual of a Planar Conformal Field
  Theory in 4D''},
\nbbstjournal{\doiref{10.1103/physrevlett.123.081602}{Physical~Review~Letters~123,~A.~Sever~(2019)}},
\nbbsteprint{\arxivref{1903.10508}{arxiv:1903.10508}},
\href{http://dx.doi.org/10.1103/PhysRevLett.123.081602}{\nbbsturl{http://dx.doi.org/10.1103/PhysRevLett.123.081602}}.

\bibitem{Gromov2_2019}
\nbbstauthor{N.~Gromov and A.~Sever},
\nbbsttitle{``Quantum fishchain in AdS5''},
\nbbstjournal{\doiref{10.1007/jhep10(2019)085}{Journal~of~High~Energy~Physics~2019,~A.~Sever~(2019)}},
\nbbsteprint{\arxivref{1907.01001}{arxiv:1907.01001}},
\href{http://dx.doi.org/10.1007/JHEP10(2019)085}{\nbbsturl{http://dx.doi.org/10.1007/JHEP10(2019)085}}.

\bibitem{Gromov_2020}
\nbbstauthor{N.~Gromov and A.~Sever},
\nbbsttitle{``The holographic dual of strongly $\gamma$-deformed $\mathcal{N}$
  = 4 SYM theory - derivation, generalization, integrability and discrete
  reparametrization symmetry''},
\nbbstjournal{\doiref{10.1007/jhep02(2020)035}{Journal~of~High~Energy~Physics~2020,~A.~Sever~(2020)}},
\nbbsteprint{\arxivref{1908.10379}{arxiv:1908.10379}},
\href{http://dx.doi.org/10.1007/JHEP02(2020)035}{\nbbsturl{http://dx.doi.org/10.1007/JHEP02(2020)035}}.

\bibitem{Fokken_2014}
\nbbstauthor{J.~Fokken, C.~Sieg and M.~Wilhelm},
\nbbsttitle{``Non-conformality of ${{\gamma }_{i} }$-deformed $\mathcal{N}$ = 4
  SYM theory''},
\nbbstjournal{\doiref{10.1088/1751-8113/47/45/455401}{Journal~of~Physics~A:~Mathematical~and~Theoretical~47,~455401~(2014)}},
\nbbsteprint{\arxivref{1308.4420}{arxiv:1308.4420}},
\href{http://dx.doi.org/10.1088/1751-8113/47/45/455401}{\nbbsturl{http://dx.doi.org/10.1088/1751-8113/47/45/455401}}.

\bibitem{Sieg_2016}
\nbbstauthor{C.~Sieg and M.~Wilhelm},
\nbbsttitle{``On a CFT limit of planar $\gamma_i$-deformed $\mathcal{N}$=4 SYM
  theory''},
\nbbstjournal{\doiref{10.1016/j.physletb.2016.03.004}{Physics~Letters~B~756,~118~(2016)}},
\nbbsteprint{\arxivref{1602.05817}{arxiv:1602.05817}},
\href{http://dx.doi.org/10.1016/j.physletb.2016.03.004}{\nbbsturl{http://dx.doi.org/10.1016/j.physletb.2016.03.004}}.

\bibitem{Grabner_2018}
\nbbstauthor{D.~Grabner, N.~Gromov, V.~Kazakov and G.~Korchemsky},
\nbbsttitle{``Strongly $\gamma$-deformed $\mathcal{N}$ Supersymmetric
  Yang-Mills Theory as an Integrable Conformal Field Theory''},
\nbbstjournal{\doiref{10.1103/physrevlett.120.111601}{Physical~Review~Letters~120,~G.~Korchemsky~(2018)}},
\nbbsteprint{\arxivref{1711.04786}{arxiv:1711.04786}},
\href{http://dx.doi.org/10.1103/PhysRevLett.120.111601}{\nbbsturl{http://dx.doi.org/10.1103/PhysRevLett.120.111601}}.

\bibitem{Hogervorst_2017}
\nbbstauthor{M.~Hogervorst, M.~Paulos and A.~Vichi},
\nbbsttitle{``The ABC (in any D) of logarithmic CFT''},
\nbbstjournal{\doiref{10.1007/jhep10(2017)201}{Journal~of~High~Energy~Physics~2017,~A.~Vichi~(2017)}},
\nbbsteprint{\arxivref{1605.03959}{arxiv:1605.03959}},
\href{http://dx.doi.org/10.1007/JHEP10(2017)201}{\nbbsturl{http://dx.doi.org/10.1007/JHEP10(2017)201}}.

\bibitem{1993log}
\nbbstauthor{V.~Gurarie},
\nbbsttitle{``Logarithmic operators in conformal field theory''},
\nbbstjournal{\doiref{10.1016/0550-3213(93)90528-w}{Nuclear~Physics~B~410,~535~(1993)}},
\nbbsteprint{\arxivref{hep-th/9303160}{hep-th/9303160}},
\href{http://dx.doi.org/10.1016/0550-3213(93)90528-W}{\nbbsturl{http://dx.doi.org/10.1016/0550-3213(93)90528-W}}.

\bibitem{he2021note}
\nbbstauthor{Y.~He and H.~Saleur},
\nbbsttitle{``A note on the identity module in $c=0$ CFTs''},
\nbbsteprint{\arxivref{2109.05050}{arxiv:2109.05050}}.

\bibitem{Minahan_2003}
\nbbstauthor{J.~A.~Minahan and K.~Zarembo},
\nbbsttitle{``The Bethe-ansatz for Script $\mathcal{N}$ = 4 super
  Yang-Mills''},
\nbbstjournal{\doiref{10.1088/1126-6708/2003/03/013}{Journal~of~High~Energy~Physics~2003,~013~(2003)}},
\nbbsteprint{\arxivref{hep-th/0212208}{hep-th/0212208}},
\href{http://dx.doi.org/10.1088/1126-6708/2003/03/013}{\nbbsturl{http://dx.doi.org/10.1088/1126-6708/2003/03/013}}.

\bibitem{Lieb:1961fr}
\nbbstauthor{E.~H.~Lieb, T.~Schultz and D.~Mattis},
\nbbsttitle{``{Two soluble models of an antiferromagnetic chain}''},
\nbbstjournal{\doiref{10.1016/0003-4916(61)90115-4}{Annals~Phys.~16,~407~(1961)}}.

\bibitem{2005sprol}
\nbbstauthor{M.~Spradlin and A.~Volovich},
\nbbsttitle{``The one-loop partition function of super-Yang-Mills theory on''},
\nbbstjournal{\doiref{10.1016/j.nuclphysb.2005.01.007}{Nuclear~Physics~B~711,~199~(2005)}},
\nbbsteprint{\arxivref{hep-th/0408178}{hep-th/0408178}},
\href{http://dx.doi.org/10.1016/j.nuclphysb.2005.01.007}{\nbbsturl{http://dx.doi.org/10.1016/j.nuclphysb.2005.01.007}}.

\bibitem{stanley_2011}
\nbbstauthor{R.~P.~Stanley},
\nbbsttitle{``Enumerative Combinatorics''},
2nd edition,
Cambridge University Press (2011).

\bibitem{Flohr:2013dma}
\nbbstauthor{M.~Flohr and M.~Koehn},
\nbbsttitle{``{What the characters of irreducible subrepresentations of Jordan
  cells can tell us about LCFT}''},
\nbbstjournal{\doiref{10.1088/1751-8113/46/49/494007}{J.~Phys.~A~46,~494007~(2013)}},
\nbbsteprint{\arxivref{1307.5844}{arxiv:1307.5844}}.

\bibitem{caetano2018chiral}
\nbbstauthor{J.~Caetano, O.~Gurdogan and V.~Kazakov},
\nbbsttitle{``Chiral limit of N = 4 SYM and ABJM and integrable Feynman
  graphs''},
\nbbsteprint{\arxivref{1612.05895}{arxiv:1612.05895}}.

\bibitem{2008min}
\nbbstauthor{J.~Minahan and K.~Zarembo},
\nbbsttitle{``The Bethe ansatz for superconformal Chern-Simons''},
\nbbstjournal{\doiref{10.1088/1126-6708/2008/09/040}{Journal~of~High~Energy~Physics~2008,~040~(2008)}},
\nbbsteprint{\arxivref{0806.3951}{arxiv:0806.3951}},
\href{http://dx.doi.org/10.1088/1126-6708/2008/09/040}{\nbbsturl{http://dx.doi.org/10.1088/1126-6708/2008/09/040}}.

\bibitem{2018QSC}
\nbbstauthor{V.~Kazakov},
\nbbsttitle{``Quantum Spectral Curve of $\gamma$-Twisted $\mathcal{N}$= 4 SYM
  Theory and Fishnet CFT''},
\nbbstjournal{\doiref{10.1142/s0129055x1840010x}{Reviews~in~Mathematical~Physics~30,~1840010~(2018)}},
\nbbsteprint{\arxivref{1802.02160}{arxiv:1802.02160}},
\href{http://dx.doi.org/10.1142/S0129055X1840010X}{\nbbsturl{http://dx.doi.org/10.1142/S0129055X1840010X}}.

\end{thebibliography}

\end{document}